\documentclass[11pt]{article} 

\usepackage[newenum]{paralist}
\usepackage{graphicx}
\usepackage{ifthen}
\usepackage{amssymb}
\usepackage{ifthen}

\newcommand{\ifPS}{\ifthenelse{1 < 2} }

\newcommand{\ignore}[1]{}

\newcommand{\cts}{{mixed }}
\newcommand{\Cts}{{Mixed }}

\newtheorem{lemma}{\noindent{\bf Lemma}}

\usepackage{color}
\usepackage{soul}

\definecolor{gld1}{rgb}{.7,.30,.0} 

\newcommand{\Col}[1]{{#1}}

\newcommand{\Cut}[1]{}

\definecolor{gold}{rgb}{.5,.45,.05} 

\newcommand{\richardCol}[1]{{#1}}

\newcommand{\RevCol}[1]{{#1}}

\newcommand{\RevCut}[1]{}

\newcommand{\richardColNew}[1]{{#1}}

\newcommand{\myCol}[1]{{#1}}

\newcommand{\mynCol}[1]{{#1}}

\newcommand{\myCut}[1]{}

\newcommand{\mynCut}[1]{}

\newcommand{\kavfive}[1]{#1}
 
\newcommand{\Alter}[2]{\Cut{#1} \Col{#2}}

\newcommand{\RemoveThis}[1]{}

 \usepackage{mathtools}
\usepackage[nolist]{acronym}
\begin{acronym}
    \acro{OFDMA}{Orthogonal Frequency-Division Multiple Access}
    \acro{NGMN} {Next Generation Mobile Network}
    \acro{WiMAX}{Worldwide Interoperability for Microwave Access}
    \acro{LTE}{Long Term Evolution}
    \acro{RAN}{Radio Access Networks}
    \acro{QoS}{Quality of Service}
    \acro{ICIC}{Inter-Cell Interference Coordination}
    \acro{PC}{Power Control}
    \acro{FFR}{Fractional Frequency Reuse}
    \acro{FL}{Fractional Load}
    \acro{BCR}{Block Call Rate}
    \acro{eNB}{enhanced eNode B}
    \acro{CDMA}{Code Division Multiple Access}
    \acro{GSM}{Global System for Mobile Communications}
    \acro{BS}{Base Station}
    \acro{PRB}{Physical Resource Block}
    \acro{NG}{Next Generation}
    \acro{B3G}{Beyond 3G}
    \acro{SON}{Self-Organizing Networks}
    \acro{3GPP}{3rd Generation Partnership Project}
    \acro{UE}{User Equipment}
    \acro{RRM}{Radio Resource Management}
    \acro{PS}{Packet Scheduling}
    \acro{KPIs}{Key Performance Indicators}
    \acro{SINR}{Signal to Interference plus Noise Ratio}
    \acro{HSDPA}{High Speed Downlink Packet Access}
    \acro{RR}{Round Robin}
    \acro{TDMA}{Time Division Multiple Access}
    \acro{MTP}{Max Throughput}
    \acro{PF}{Proportional Fair}
    \acro{MMF}{Max-Min Fair}
    \acro{MAB}{Multi-Armed-Bandit}
    \acro{RB}{Restless Bandit}
    \acro{ODE}{Ordinary Differential Equation}
    \acro{i.i.d}{independent and identically distributed}
    \acro{MIMO}{Multiple Input Multiple Output}
    \acro{c.d.f}{cumulative distribution function}
    \acro{p.d.f}{probability density function}
    \acro{PPRBS}{Per Physical Resource Block Scheduler}
    \acro{KPI}{Key Performance Indicator}
    \acro{SIC}{Successive Interference Cancellation}
    \acro{FTP}{File Transfer Protocol}
		\acro{PSO}{Particle Swarm Optimization}
    \acro{FTT}{File Transfer Time}
    \acro{BS}{Base Station}
    \acro{a.s}{almost surely}
    \acro{r.v}{random variable}
    \acro{AWGN}{Additive White Gaussian Noise}
\end{acronym}

\newcommand{\ZZ}{\mathbb{Z}}
\newcommand{\RR}{\mathbb{R}}
\newcommand{\NN}{\mathbb{N}}

\newcommand{\tends}[2]{\underset{#1 \to #2}{\to}}

\DeclarePairedDelimiter{\norm}{\lVert}{\rVert}

\newcommand{\expec}[1]{E \left[ #1 \right] }

\newcommand{\limu}[2]{ \underset{#1 \to #2} {\lim} \,}

\newcommand{\indic}{1}
\newtheorem{thm}{Theorem}

\begin{document}
\title{\Cts Polling  with Rerouting and Applications} 
\author{Veeraruna Kavitha \and Richard Combes}
\date{\today}
\maketitle
\begin{abstract}
Queueing systems with a single server \richardColNew{in which} customers wait to be served at a finite number of \myCol{distinct} locations (buffers/queues) are called discrete polling systems. Polling systems in which arrivals of users occur \mynCol{anywhere} in a continuum are called continuous polling systems. \Col{Often one  encounters a combination of the two systems: the users can either arrive in a continuum or wait in a finite set (i.e. wait at a finite number of queues).
We call these systems \cts polling systems.}
Also, in some applications, customers are rerouted to a new location (for another service) after their service is completed. 
In this work, we study \cts polling systems with rerouting. We obtain their steady state performance by discretization using the known pseudo conservation laws of discrete polling systems.
Their stationary expected workload is obtained as a limit of the stationary expected workload of a discrete system.  
The main tools for our analysis are: a) the fixed point analysis of infinite dimensional operators and; b) the convergence of Riemann sums to an integral. 
 
We analyze two applications using our results on \cts polling systems and discuss the optimal system design. We consider a local area network, in which a moving ferry facilitates communication (data transfer) using a wireless link. We also consider a distributed waste collection system \richardColNew{and derive} the optimal collection point. \Col{In both examples, the service requests can arrive anywhere in a subset of the two dimensional plane. Namely, some users arrive in a continuous set  while others wait for their service in a finite set.
The only polling systems that can model these applications are \cts systems with rerouting \richardColNew{as} introduced in this manuscript. \footnote{Parts of this work were carried out while the first author was at INRIA, France. Parts of this work were carried out while the second author was at Orange Labs and INRIA, France.} }
\end{abstract}

\section{Introduction}
Polling systems are a special class of queueing systems where a single server \richardCol{serves} more than one queue.
A certain amount of time is required to switch \richardColNew{from one queue to the next}. \richardCol{During this time the server does not serve any user and is considered idle. Hence polling systems are non-work conserving queueing systems.} 
 
  \richardCol{A queue usually represents a location where users form a waiting line and wait to be served. In this case, the idle/switching time is the time required for the server to travel between these locations.} Various types of polling systems are studied in the literature. \Col{These} \richardCol{systems usually differ in: the order in which queues are served, the service discipline for each queue or the number of queues (etc)}. \richardColNew{We will mention a few possible models.}

A system with a finite number of queues is called a discrete polling system. This is the most studied model in the literature. 
\richardCol{In this paper we also consider arrivals that can occur anywhere in a continuum, as in \cite{JournPaper,Coffman1,Eliazar}}. Waiting lines are not formed for arrivals in a continuum, since the probability of \richardCol{observing} two users waiting at the same point is zero if the arrival measure is continuous. We
use the word ``queue'' to refer to any location where a user can wait for service, irrespective of whether or not a waiting line is formed. We first discuss the discrete models \richardCol{and then} the models with two types of arrivals, which we call mixed systems.
 
  The queues are usually served in a deterministic cyclic order. They can also be served in a random order.
  The server \richardCol{usually moves} in a fixed direction \richardCol{along a path}. There are exceptions, for example in elevator polling the server reverses its direction after completing a cycle (\cite{elevatorpoll}). 
  There are various possible \Col{service} disciplines, \richardCol{namely}:
\begin{enumerate}
 \item \emph{globally gated service} (e.g., \cite{Boxma}): a global gate is closed for all the queues when the server reaches a given queue, say queue $1$.  
The time between two successive visits to queue $1$ is called a cycle. 
Only the users that arrived before the gate closure are served, \richardColNew{as the server goes through the entire cycle}. \richardCol{Users that arrived after the gate closure are served during the next cycle.}
\item \emph{gated service} (e.g \cite{VindoPoll,Boxma,Kroese4}): there is one gate per queue. When the server reaches a queue, it closes the corresponding gate, and serves the users who arrived at that queue before the gate closure. \richardCol{The server then switches to the next queue.} \Cut{The server then leaves the queue, and users who arrived at that queue after the gate closure will be served during the next cycle. }
\item \emph{exhaustive service}: (e.g., \cite{VindoPoll,Boxma,JournPaper}) \richardCol{When it arrives at a queue, the server serves the users of this queue until it is empty.} 
 \Cut{when the server reaches a queue, it starts serving the users for that queue. The server does not leave 
the queue until it is empty. }
\item \emph{mixed service systems} (e.g., \cite{Boxma2,SLF}): \richardCol{different queues have different types of services.} For instance, some queues might be served using a gated service, while the others are served using an exhaustive service.
\end{enumerate}

\mynCol{Normally} users leave the system immediately after receiving their service. A variant is polling systems with 'rerouting' (\cite{SLF}). When rerouting applies, users \richardColNew{who} have received their service can either leave the system or be transferred to a new queue to receive a new service with a given probability. \richardColNew{It is noted that the queue which the user is rerouted to can be the same as the queue where he previously received service}. It is also noted that a user might be rerouted multiple times.

\myCol{As already mentioned polling} systems with a finite number of queues are called discrete polling systems.  
{\it Continuous polling systems are polling systems in which \richardCol{users wait for service anywhere in a continuous set (a subset of a plane).}} A server moves along a closed path, stops when it encounters a user requiring service and resumes its journey after service completion. 

\richardCol{The above variants of discrete polling systems can be extended to continuous polling systems.}  In {\it gated service}, users are served immediately upon encountering the server. Exhaustive service and gated service are equivalent when there are no waiting lines.  In \cts  polling systems 
\RevCol{(waiting lines can be formed due to discrete arrivals)}
with exhaustive service the server leaves a queue only when it is empty. In {\it globally gated service} (\cite{JournPaper}), the server closes a global gate when it reaches a \Col{pre-}determined point (say $0$) and marks all the users that arrived before the gate closure. The server goes through its path and only serves the marked users before returning to $0$. Most continuous polling systems are studied under the standard {\it gated service} assumption (\cite{Fuhrmann,Coffman1,Coffman2,Kroese4,Kroese1,Kroese2,Kroese3,Eliazar,Eliazar1}). As in the discrete case, there can be rerouting. Users who have received their service are transfered to a different location for another service \richardCol{according to} some probability distribution. In \cite{Value}, a continuous polling system \richardCol{is considered, where} users can be rerouted only one time. In this paper we extend it to an arbitrary finite number of reroutings.

\medskip
\medskip

\richardCol{
We consider a random number of reroutings. The number of reroutings is upper bounded by a finite constant with probability one. The extension to an infinite number of reroutings would be a non-trivial extension. We assume Poisson arrivals. This assumption is used extensively in the proofs by application of Wald's lemma. \richardColNew{Similarly, any other generalization of assumptions would be non-trivial.}}
 
 \noindent{\it Our contribution:} 

The main contribution of this paper is to introduce a polling model which bridges the gap between discrete and continuous polling systems and generalizes them. We call such a model a {\it \cts polling system}. \richardCol{We call the distribution of the location of users arrivals the ``arrival position distribution''.} In a discrete polling system, the arrival position distribution is a discrete distribution on a finite set. In a continuous polling system, the arrival position distribution has a density with respect to the Lebesgue measure. In other words the distribution has no atoms \richardCol{and} the probability of arriving at any point is $0$. A \cts polling system combines the two: the arrival position distribution is a sum of a discrete distribution with a finite number of atoms, and a continuous distribution which has a density with respect to the Lebesgue measure.
  
	Extending existing results \richardCol{on} continuous and discrete polling systems to \cts polling systems requires significant changes in the proof techniques with respect to previous papers (\cite{Value,JournPaper}) \Col{and these issues are discussed later}.
	\Col{Following a discretization approach\footnote{
\Col{The proof in~\cite{Value} has an error, especially in Lemmas 3 and 4. This error is corrected in \cite{JournPaper} and we are  
using  mainly the techniques of \cite{JournPaper} here, adapting \richardCol{them} to rerouting and to mixed arrivals.}} \richardCol{as in} the earlier papers~\cite{JournPaper,Value}, we obtain the stationary expected workload as the limit of the stationary expected workload of a discrete system. We discretize the \cts polling system in such a way that the pseudo conservation laws of Sidi et al. (\cite{SLF}) for discrete polling systems with rerouting are applicable.
	}
	
	\Cts polling systems significantly extend the discrete and continuous polling models. We show in this paper that even for a fairly simple \Col{application}, the arrival position distribution has both continuous and discrete components. Hence \cts polling systems \richardColNew{enable the analysis of} many \richardCol{new} applications of interest. 

To the best of our knowledge this is the first work to consider a polling system \richardCol{which}:
\begin{itemize}
\item has an arbitrary finite number of reroutings, and
\item generalizes continuous and discrete polling systems i.e. a \cts polling system.
\end{itemize}

\subsection*{Applications}
In a Ferry assisted Wireless LAN (FWLAN), nodes are not connected directly and use a mobile relay known as a message ferry. The message ferry serves as a ``postman'' and delivers messages exchanged by the nodes. Nodes might be static or dynamic. The ferry moves periodically \richardCol{on a} cyclic route and stops when it is in the vicinity of a user with data to transmit or receive. The ferry must be close enough \richardCol{to the user} to ensure reliable communication over a wireless link. \richardCol{It is noted that the network area can be large and the transmission power can be limited}. 

	We study \richardCol{a} FWLAN, where both local and external data transfers occur. 
{Namely, there is a base station which is a gateway to the external world. 
During an external data transfer, the ferry receives the message from the base station, and delivers \richardCol{it} to the sink, when it reaches the vicinity of the sink (and vice-versa).
For \richardCol{a} local data transfer the source and \richardCol{the} sink are arbitrary points on the path. \richardCol{The only difference with an external transfer is that the ferry receives the message from a source user (located anywhere in the area)}.  Finally we have a hybrid data transfer when a local data transfer is accomplished with the help of the base station. External and local data transfers require \richardCol{two} services at two locations while hybrid transfers \Col{require service} at three locations. To model such FWLANs we need polling systems that support more than one service such as the new polling systems studied in this paper.
}

 As far as the network geometry is concerned, we consider rectangular ferry paths. In this model the arrival position distribution 
\Col{(when ``projected'' on the rectangular ferry path) } has atoms at the corner points, and has a density with respect to the Lebesgue measure elsewhere.
 \Col{This cannot be modeled either by a \richardColNew{continuous polling system} or by a discrete polling system. It actually requires the combination of the two, i.e. the \cts polling systems studied in this paper to analyze \Col{the system performance}.}
 
 We consider a second application which is a distributed waste collection system. A \richardCol{van} collects waste from locations that are spread over a large geographical area. Waste generation occurs at a stationary rate, for example when rescue operations take place in an area hit by a natural disaster or in a production unit where production of intermediate products takes place at \richardCol{different} geographical locations. A FWLAN where information is collected from all the nodes and transmitted to a fixed gateway also falls under this category. {\it Continuous models are required whenever the waste generation points are arbitrary and vary from cycle to cycle. Even if the waste generation points are fixed but are large in number, one can obtain simpler formulas by \richardCol{approximating} the system \richardCol{by} a continuous polling model.} Given the distribution of the waste generation points and the rate of generation, we obtain a waste collection point \richardCol{minimizing} the stationary workload.  Here again one requires the \cts polling systems of this paper, because the first service (waste collection) occurs anywhere in the path, while the second service (\richardCol{discarding} the waste) happens at one fixed point.
  
 In all applications, we \richardCol{minimize} the stationary average virtual workload. This is Pareto optimal for all the waiting users optimizing their expected waiting times (see \cite{JournPaper}).


\subsection*{Related Work}
Continuous polling systems were first introduced by Fuhrmann and Cooper \cite{Fuhrmann}, further explored by Coffman and Gilbert \cite{Coffman1,Coffman2} and  Kroese and Schmidt \cite{Kroese4,Kroese1,Kroese2,Kroese3} in a series of works. Stability results are available in \cite{space,spa,liu}.
The continuous systems are usually analyzed under simplified conditions, which we \richardCol{call} ``symmetric conditions'': arrival instants form a Poisson process and the arrival position distribution is uniform on a circle. The server moves at a constant speed in a fixed direction. \richardCol{The} service times are identically distributed.

Snowplowing systems generalize many of the above assumptions. For example, the incoming \richardCol{work} to the system is taken to be a general L\'evy random measure and the walking times are assumed to be random (see for example \cite{Eliazar,Eliazar1}).

In the literature, continuous polling systems have \richardCol{usually} been analyzed with simple gated/exhaustive service.
In \cite{JournPaper} \mynCol{one of the authors} study continuous polling systems \richardCol{with}: globally gated service or gated service or a mixture of \richardCol{both} or elevator service, without requiring symmetry. 

In this paper we do not use the symmetric assumptions and \richardCol{consider rerouting as well as \cts systems. Rerouting and mixed position arrival measure  make this paper significantly different from \Col{the} previous papers (\cite{JournPaper,Value}) in terms of results, applications and proof techniques required for obtaining the stationary workload.}

Polling systems are widely used for modeling and analyzing communication systems, computer hardware and software, and road traffic control. For example, in \cite{WSaad,itc,wiopt}, authors analyze FWLANs. Apart from  FWLANs, there are many other applications in which customers may require more than one service.   
Such applications are mentioned in \cite{SLF}, for instance:
1) Automatic Repeat reQuest (ARQ) protocols used for \richardCol{transmission error recovery};
2) Multi threading computers where a single processor \richardCol{performs} multiple tasks: computation, disk and memory I/O, switching between threads etc.

%
\RevCut{
The results of this paper are useful in the aforementioned applications, whenever the arrivals are on a continuum and when there are finite number of reroutings}.  
 
The rest of the paper is organized as follows: the system model and notations  are introduced in section \ref{Sec_CtsPoll}. In the same section, we  state the main result of the paper, Theorem \ref{Thrm_MixedService}.
We discuss the application \richardCol{of our results} to FWLANs and a distributed waste collector in sections \ref{section_fwlan} and \ref{section_collect} respectively. The proof of Theorem 1 is given in section~\ref{section_proof}.
The appendices contain some details of the proofs. Appendix D summarizes the main notations used in this article.

\section{\Cts Polling Systems}
\label{Sec_CtsPoll}

\Col{
A server is moving continuously \Col{in one direction} on a circle  ${\cal C}$ 
with speed $\alpha$. It stops only when it encounters a user with \Col{a} request and resumes its journey after serving the encountered user. 
We consider a \cts polling system, i.e. the users can  either arrive anywhere  on the circle ${\cal C}$ or can form 
 waiting queue(s) at \Col{a finite set of points} on ${\cal C}$. 
}

\Col{
 The external arrival process is modeled by a Poisson process with intensity $\lambda$ and every external/new arrival is associated with two marks: the position ${Q_0} \in {\cal C}$ distributed as $P_{Q_0}$ and the service time $B_{Q_0}$.
 The service time can depend upon the position ${Q_0}$ of the \richardColNew{new} arrival.
 Let $b_{0}(q)$ and $b_{0}^{(2)} (q)$ denote the first and second moments of the service time $B_{Q_0}$ conditioned on the event that the position of arrival is $q$. The service times of different users are independent of each other.
}

A user is served the first time the server encounters it on the circle. After \richardCol{his} service is completed, the user is either
 rerouted, independently of all the previous events, to a new point $q' \in {\cal C}$ with  probability $\epsilon_1 P_{Q_1}(dq')$ or exits the system with
 probability $1 - \epsilon_1.$  {\it It is noted that the rerouting probabilities are independent of the position at which the previous service is obtained.}
The { rerouted users are served the next time the server encounters them. Once again they either exit the system with probability $1-\epsilon_2$ or are rerouted to another point with probability $\epsilon_2 P_{Q_2} (dq)$ independently of all the earlier events \richardCol{and so on}. There can be at most $N$ reroutings with
$N < \infty$. The rerouting probabilities are $\{\epsilon_j P_{Q_j} (dq) \}_{0 \le j < N}$} with  $\epsilon_0 = 1$.
The arrival and rerouting probabilities can be mixed. Namely, there are positive functions $\{f_{Q_j}\}$ defined on $[0,|{\cal C}|]$ and constants $M_j < \infty$ and \Col{a finite set of} \Col{points } $\{q_{j,i}\}$ such that the probability distribution of the $j^{\mbox{th}}$ \mynCol{(rerouted)}
\richardCol{arrival is}:
\begin{eqnarray}
\label{Eqn_Arrivalprobs}
P_{Q_j} (dq) = \left ( \sum_{i=1}^{M_j} p_{j,i} 1_{\{q_{j,i}\}} (q) + f_{Q_j} (q) \right ) dq  \ \ \ \ \ \ \mbox{ for  all }  0 \le j < N.
\end{eqnarray}
\richardCol{There can be more than one user waiting at the same point with  positive probability \Col{(e.g.,  at one of the  $q_{i,j}$)}}. The server always serves the user with the largest number of services completed \footnote{\RevCol{The assumption appears like a natural routine followed in the queues: when a customer has been partially served, he is given the priority over new customers.   We believe one can probably modify the proof to accommodate other service \richardCol{disciplines}.}}. For instance, a user that has been rerouted once will be served before a user waiting for his first service if they are both waiting at the same point. The users with the same number of services completed are served in FIFO (First In First Out) order.

\richardCol{
 The users are rerouted independently of the previous events. However their service requirement for the $(j+1)$-th service depends
 on the position $Q_j$ to which they have been rerouted (after completing the $j$-th service), which is distributed as $P_{Q_j}$.} Namely the service depends only upon the position at which it is \richardCol{received}, irrespective of the number of reroutings already received.
 Let $b_{j} (q)$, $b_{j}^{(2)}(q)$ \richardCol{denote} the conditional first and second moments of the service time $B_{Q_j}$ for the $j$-rerouted users if they have been rerouted to location $q$.
Let ${\bar b}_{j}:= E_{Q_j}[b_{j} (Q_j)] $  (the expectations are with respect to $P_{Q_j}$), for every $0\le j <N$, represent the
unconditional moments. Similarly we define the second moments:  ${\bar b}_{j}^{(2)}:= E_{Q_j}[b_{j}^{(2)}({Q_j})]$.

\noindent\underline{\bf Notations:} 
The circular path is mapped to an interval  $[0, |{\cal C}|]$
($|{\cal C}|$ is the length of ${\cal C}$) \richardCol{to ease the analysis}. 
Variables \richardCol{such as} $b_{j}$,  $f_{Q_j}$, $\tau$  etc. \richardCol{denote} nonnegative functions on \richardCol{the} interval $[0, |{\cal C}|]$
while terms \richardCol{such as} $b_{j}(q)$ or $\tau(q)$ \richardCol{denote} their value at point $q \in [0, |{\cal C}|].$ \richardCol{Overline} variables, such as ${\bar b}_{j}$, \richardCol{denote their} average w.r.t. the corresponding position distribution.  
 \richardCol{The mathematical expectation is denoted by $E$.} Expectation is either with respect to $Q_j$ for \richardCol{a given} $j$ or the stationary measure of the process under consideration. $E^0$ \richardCol{denotes} the expectation with respect to the Palm stationary measure.
 To avoid ambiguity, we suffix $E$ with variables like $Q_j$. 
 Define the following (with $\epsilon_0 :=1$ ${\hat \epsilon}_j := \Pi_{i=0}^j \epsilon_i$):
 {
\begin{eqnarray*}
\rho_{j} ([a,c])   &:=& \lambda \int_a^c b_{j} (q) P_{Q_j} (dq)  ,  \ \   \rho_{j} :=  \rho_{j} ([0, |{\cal C}|]),  \mbox{ for all } 0 \le j < N\\
\rho ([a,c])   &:=&   \rho_{0}([a,c]) + \sum_{j=1}^{N-1} {\hat \epsilon}_j \rho_{j} ([a,c])   \mbox{ and }  
\rho =  \rho_{0} + \sum_{j=1}^{N-1} {\hat \epsilon}_j \rho_{j}.
\end{eqnarray*} }
\mynCol{
In general for any measurable subset $I \subset [0, |{\cal C}|]$ we extend the definition of $\rho$ in a natural way:
$$
\rho_{j} ( I)   := \lambda \int_0^{|{\cal C}|} b_{j} (q)  \indic_{\{I\} }(q)   P_{Q_j} (dq) $$
}

\subsection{Main Result}
The {\it virtual workload} of a polling system is defined as the total workload corresponding to all the waiting users,
i.e. the sum of the service times of all the waiting users. 
 Not much theory is available for calculating the expected virtual workload of polling systems with arrivals in a continuum. 
In this section we derive new (stationary expected workload) results \richardCol{for} \cts polling systems with rerouting.
  Throughout we consider
{\it stationary and ergodic systems.}
We obtain the stationary expected virtual workload:
\begin{thm}
 \label{Thrm_MixedService}
 Assume  that $\{ b_{j} ;  j \} $  and $\{ f_{Q_j} ; j\} $  are continuous functions.  
There exists a threshold $\bar{\pi}_b^{(2)} > 0$ such that if the second moments of the service times verify $ {{\bar b}_{j}^{(2)}} <
\bar{\pi}_b^{(2)}$ for all $j$, then  
 the expected stationary virtual workload  for the  \cts polling system with rerouting, $V_{rrt}$,
 equals (with ${\hat \rho} (q) := \rho ([0, q))$,  $\check {\epsilon}_j^k := \Pi_{i=j}^{k} \epsilon_i$):
 {\small
\begin{eqnarray}
 V_{rrt} 
&=&  \frac{\rho \lambda}{2(1-\rho)}  \sum_{k=0}^{N-1} \hat{\epsilon}_k {\bar b}_{k}^{(2)} +  \frac{\rho \lambda}{1-\rho}\sum_{l=0}^{N-1} \sum_{k=l+1}^{N-1} \hat{\epsilon}_l 
{\bar b}_{k} {\bar b}_{l} 
+   \frac{\rho  |{\cal C}|  \alpha^{-1}   } {2 }
\nonumber \\
&& \hspace{-10mm}
+\frac{\lambda \alpha^{-1}}{1 - \rho }  \int_0^{|{\cal C}|} \int_0^{|{\cal C}|} \left (b_{0}(q)  + \sum_{j=1}^{N-1} {\hat \epsilon}_j {\bar b}_{j} \right )
\left (  {\hat \rho} (y) -  {\hat \rho}(q)
+   {1_{\{ y < q \} } \rho}\right )  P_{Q_0}(dq)     dy   \nonumber \\
&& \hspace{-10mm}
+ \sum_{j=0}^{N-2} \frac{{\hat \epsilon}_{j+1} \lambda \alpha^{-1}}{1-\rho} \int_0^{|{\cal C}|} \int_0^{|{\cal C}|}
\left  ( b_{{j+1}} (q') +  \hspace{-2mm} \sum_{k=j+2}^{N-1} \check{\epsilon}_{j+2}^k {\bar b}_{k} \right )  (q' - q + |{\cal C}| 1_{\{q > q'\}} ) 
\nonumber \\
&& \hspace{75mm}
P_{Q_{j+1}} (dq')   P_{Q_{j}}(dq) .   \hspace{5mm} 
 \label{Eqn_WorkloadMxS}
 \label{Eqn_Vmix}
\label{Eqn_VmixLim}
 \label{Eqn_MainFormula}
\end{eqnarray}  }
\end{thm}\myCol{
{\bf Proof:} The proof is presented in section \ref{section_proof}. \hfill{$\Box$}}

\subsection{Special Cases}\label{subsec:special_cases}
In this section we compare \richardCol{our result with special cases studied in the literature so far}. We first notice that, when $N = 2$ and when the distributions $\{P_{Q_j}\}$ have a density with respect to the Lebesgue measure (they have no atoms), then the workload expression (\ref{Eqn_Vmix}) 
\richardCol{is equal to} the stationary expected workload derived in \cite[Theorem 1]{Value}. 
We now simplify this formula for some special cases.

1) Under symmetric conditions:  We have uniform arrivals so that $\{ P_{Q_j}\}$ are all uniform.
For every $j$, the service time moments $ b_{j}(q)$, $b_{j}^{(2)} (q)$ are equal at all points $q$. \richardCol{We denote them by} ${\bar b}_{j}$, ${\bar b}_{j}^{(2)}$.
Then   ${\hat \rho}(q)  =  \rho q/|{\cal C}|$ and   (\ref{Eqn_Vmix}) simplifies to:
{
 \begin{eqnarray}
V_{rrt}^{sym}  &=&
\frac{\rho \lambda}{2(1-\rho)}  \sum_{k=0}^{N-1} \hat{\epsilon}_k {\bar b}_{k}^{(2)} +  \frac{\rho \lambda}{1-\rho}\sum_{l=0}^{N-1} \sum_{k=l+1}^{N-1} \hat{\epsilon}_l 
{\bar b}_{k} {\bar b}_{l} 
+   \frac{\rho {|{\cal C}|}   \alpha^{-1}   } {2 (1-\rho)}  \nonumber \\
&&+ \sum_{j=0}^{N-2} \frac{{\hat \epsilon}_{j+1} \lambda \alpha^{-1}}{1-\rho} 
\left  ( {\bar b}_{{j+1}}  +  \hspace{-2mm} \sum_{k=j+2}^{N-1} \check{\epsilon}_{j+2}^k {\bar b}_{k} \right )  \frac{|{\cal C}|}{2}.
\label{Eqn_Vmixsym}
 \end{eqnarray}
}

\RemoveThis{
1b) Under symmetric conditions with atoms: 
Say $P_{Q_j}$ for all $j$ equals (with $p^c:= (1 - \sum_{i} p_i)/{|{\cal C}|}$):
$$
P_{Q_j} (dq) =     p^c dq + \sum_{i=1}^M p_i 1_{\{q_{i} \}} (q) dq.
$$
Then,
$$
{\hat \rho} (q) = (1 + \sum_j {\hat \epsilon}_j ) {\bar b} \left ( {p}^c q + \sum_{i=1}^M p_i 1_{\{q_{i} < q \}}  \right ) 
$$
\begin{eqnarray}
V_{rrt}^{sym,cross}  &=&
\frac{\rho \lambda}{2(1-\rho)}  \sum_{k=0}^{N-1} \hat{\epsilon}_k {\bar b}_{k}^{(2)} +  \frac{\rho \lambda}{1-\rho}\sum_{l=0}^{N-1} \sum_{k=l+1}^{N-1} \hat{\epsilon}_l 
{\bar b}_{k} {\bar b}_{l} 
+   \frac{\rho {|{\cal C}|}   \alpha^{-1}   } {2 (1-\rho)}  \nonumber \\
&&+ \sum_{j=0}^{N-2} \frac{{\hat \epsilon}_{j+1} \lambda \alpha^{-1}}{1-\rho} 
\left  ( {\bar b}_{{j+1}}  +  \hspace{-2mm} \sum_{k=j+2}^{N-1} \check{\epsilon}_{j+2}^k {\bar b}_{k} \right )  \frac{|{\cal C}|}{2}.
\label{Eqn_Vmixsymcross}
 \end{eqnarray}
}

2) Gated  polling   under symmetric conditions:  (similar to case 1) is analyzed  in \cite{Kroese4}.  By \cite[Theorem 5.1]{Kroese4},  the stationary expected number of waiting users is:
\begin{eqnarray*}
E[L] = \lambda {\bar b}_{0}+ \frac{\lambda\left(  \alpha^{-1} + \lambda {\bar b}_{0}^{(2)} \right)}{2\left(1-\lambda {\bar b}_{0}\right)} 
\mbox{ with }  |{\cal C}|=1.
\end{eqnarray*}
The previous expression also includes the user under service.
Excluding the user under service  {(i.e. excluding the term $\lambda {\bar b}_0$)} and applying Wald's lemma the expected virtual workload  equals :
\[
V^{sym}_g =        \frac{\lambda {\bar b}_{0}\left(  \alpha^{-1} + \lambda {\bar b}_{0}^{(2)} \right)}{2\left(1-\lambda {\bar b}_{0}\right)}.
\]
This matches with (\ref{Eqn_Vmixsym}) when $|{\cal C}| = 1$,  $N=1$  (no rerouting).

3) Gated polling system under general conditions:  In \cite{JournPaper}, \mynCol{a mixed service polling system  is considered,} \richardCol{with both
gated and globally gated service}. From~\cite[Theorem 1]{JournPaper} the expected stationary virtual 
workload for pure gated service is obtained by substituting   $p_{gg} = 0 = 1-p_g$ \myCol{in \cite[Theorem 1]{JournPaper}} and it equals:
\begin{eqnarray}
\label{Eqn_Ve}
 V_{g}
 &=&    \lambda {\bar b}\frac{ \lambda {\bar b}^{(2)}} {2 (1-\lambda {\bar b})}  +
 \frac{\lambda {\bar b}|{\cal C }| \alpha^{-1}    } {2 }
+ \frac{|{\cal C}| \alpha^{-1}} {2(1-\lambda {\bar b})}
\left ( \lambda^2 {\bar b}^2  \right ).
 \end{eqnarray}
 The first integral in \richardCol{formula} (\ref{Eqn_Vmix}) with no rerouting ($N=1$)  equals\footnote{By \richardCol{exchanging} the order of the two integrals,
\begin{eqnarray*}
E [b_{0}({Q_0}) {\hat b}_{0}({Q_0}) ] = \int_0^{|{\cal C}|}\left ( \int_0^{q'}  b_{0}(q)  P_{Q_0}(dq)     \right ) b_{0}(q') P_{Q_0}(dq')   \hspace{-56mm}\\
&  & \hspace{-22mm} =
\int_0^{|{\cal C}|} \left ( \int_q^ {|{\cal C}|}  b_{0}(q') P_{Q_0}(dq')    \right )   b_{0}(q) P_{Q_0}(dq) \\
&  & \hspace{-22mm}
=  \int_0^{|{\cal C}|} \left ( {\bar b}_{0} -  \int_0^q  b_{0}(q')  P_{Q_0}(dq')    \right ) b_{0}(q) P_{Q_0}(dq)
\hspace{3mm}
\mbox{ 
and so,  $E [b_{0}({Q_0}) {\hat b}_{0}({Q_0}) ]  = {\bar b}^2_{Q_0}/2$. }\end{eqnarray*}
}:
\begin{eqnarray*}
\frac{\lambda \alpha^{-1}}{1 - \rho } \lambda \int_0^{|{\cal C}|} \int_0^{|{\cal C}|}
 b_{0}(q')
\left (  {\hat b}_{0} (q) -  {\hat b}_{0} (q')  +   {1_{\{ q < q' \} } {\bar b}_{0} }\right )  P_{Q_0}(dq')    dq  \hspace{-75mm} \\
&& \hspace{-28mm}= \frac{\lambda \alpha^{-1}}{1 - \rho } \lambda \left ({\bar b}_{0} \hspace{-1mm} \int_0^{|\cal C|}\hspace{-3mm} {\hat b}_{0} (q) dq - |{\cal C}| E_{Q_0} [b_{0}({Q_0}) {\hat b}_{0} ({Q_0})]
   + {\bar b}_{0}  \hspace{-1mm} \int_0^{|\cal C|}\hspace{-3mm} \left ( {\bar b}_{0} - {\hat b}_{0} (q) \right ) dq  \right ) \\
&&\hspace{-20mm}= \frac{\lambda \alpha^{-1}}{1 - \rho } \lambda  {\bar b}_{0}^2 \frac{|{\cal C}| }{2}, \ \ \mbox{ where } {\hat b}_{0} (q) :=  \int_{[0,q)} b_{0}(y) P_{Q_0}(dy). 
\end{eqnarray*}
Upon further simplification,  (\ref{Eqn_Vmix})  (with $N=1$) matches with  (\ref{Eqn_Ve}).

4) Globally gated systems under general conditions:  A  globally gated system can be obtained from our \cts polling system with rerouting by substituting 
$N=2$, $P_{Q_0} (dq) = 1_{\{q = 0\}}$,  $b_{0} (0) = 0$,  $\epsilon_1 = 1$. \richardCol{Then $P_{Q_1}$ denotes the \Alter{actual arrival position distribution}{distribution of the position at which  the user is waiting for service}  and $B_{Q_1}$
\Alter{the actual service requirements}{denotes the service time of the user}. All the arrivals \Col{occur} at point $0$ and are rerouted to the positions where (actual) service is \Col{received} only after the server reaches the point $0$. \Col{Zero service time for the first service ensures that users are immediately rerouted from point $0$ to mimic the globally gated service discipline}.} In this case (\ref{Eqn_Vmix}) simplifies 
to\footnote{It is noted that, since $P_{Q_0}$  is concentrated only at $0$,  the first integral of (\ref{Eqn_Vmix}) simplifies to: 
$$\frac{\lambda \alpha^{-1} }{1-\rho } \int_0^{|{\cal C}|} ({\bar b}_{1})  (\hat{\rho} (y)  ) dy  
\, \, \, = \, \, \, \frac{\lambda  \alpha^{-1} }{1-\rho } \lambda {\bar b}_{1} \left (|{\cal C}|  {\bar b}_{1} - E [Q_1 b_{1}(Q_1)] \right ).$$
}:
\begin{eqnarray*}
V_{gg} = \frac{\rho \lambda {\bar b}_{1}^{(2)} }{2(1-\rho)} 
+ \frac{\rho |{\cal C}| \alpha^{-1} }{2}  \frac{1+\rho}{1-\rho} +  \lambda \alpha^{-1}  E_{Q_1}[Q_1 b_{1}(Q_1)],
\end{eqnarray*} which matches with the formula  derived for globally gated system in \cite[eqn. (5)]{JournPaper}.

\Col{In the next two sections we consider two applications which can only be modeled by the polling models of this paper and illustrate the use of formula  (\ref{Eqn_Vmix}). The proof of Theorem 1 is \richardCol{presented in} section \ref{section_proof}.
}

\section{\Col{Application:} Ferry based Wireless LAN}
\label{section_fwlan}

\begin{figure}[!htbp]
\includegraphics[width=0.7\columnwidth]{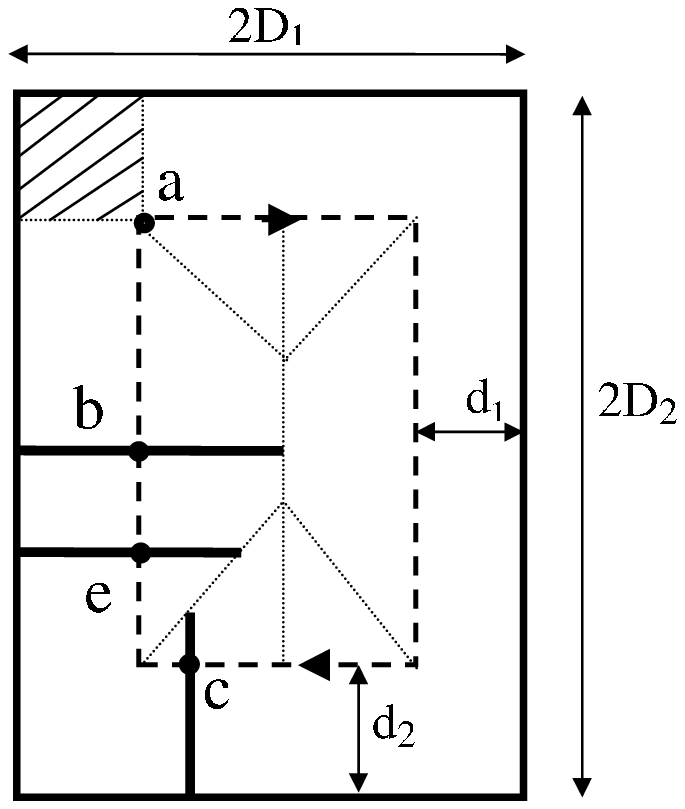} 
\caption{ Ferry in a rectangular area \label{Fig_RectFWLAN}}
\end{figure}

 \richardCol{We consider a local area network  in a rectangular area $\Delta$ of length $2D_1$ and breadth $2D_2$ (see Figure \ref{Fig_RectFWLAN}).}
There is no direct connectivity between users, and communication requires a mobile relay called a ``ferry''. The ferry moves along a closed rectangular
path ${\cal C} \subset \Delta$, with constant speed $\alpha$.  As will be shown shortly, this is a configuration that cannot be analyzed using any of the previous results. For other configurations of FWLANs the reader can refer to previous papers (\cite{Value,JournPaper}). 

We call local data transfer a data transfer between two users. For a local data transfer, the ferry collects the data from the source user when it encounters \richardCol{him}. We call this {\it uplink service.}
The uplink data comes with the address of the destination user to which it is intended. After the ferry has reached the destination user, it delivers the data. We call this  {\it downlink service.} Hence every local data transfer requires two services.
 We call external data transfer a data transfer between a user and the external world. There is a base station (BS) which serves as the gateway to the external world. For an external data transfer, the ferry collects data from a user and delivers it to the BS (or vice-versa). \Col{So external data transfer also requires two services. Here \Col{the} BS becomes one of the two users and \Col{the rest is identical to a local data transfer}.} In the next subsection, we also consider a hybrid configuration, where the BS plays a role in the data transfer between users.



{\it Optimization:}
The ferry path ${\cal C}_{d_1,d_2}$ is completely determined by $d_1$ and $d_2$, the distances of the ferry path from the sides of the rectangular area as \richardCol{depicted} in Figure \ref{Fig_RectFWLAN}.
 Let point $a$ \richardCol{correspond} to point $0$ of the line segment representing the entire cycle path.
We consider a suboptimal problem with $d_1 = d_2$. This will simplify the explanations and result in a one dimensional optimization problem. One can easily extend the computations below to the case where $d_1 \ne d_2$.  {\it Let $d = d_1 = d_2$. We denote the ferry path by ${\cal C}_d$. Our goal is to obtain the optimal path, ${\cal C}_{d^*_V}$. }
As discussed in the introduction, we are interested in minimizing the expected virtual workload (\richardCol{given by} Theorem 1). {\it We obtain the expected virtual workload by modeling the FWLAN as a polling system.} \richardCol{We omit some modeling details which are similar to~\cite{JournPaper} and concentrate only on the modeling details specific to the FWLAN considered in this paper.}
Since every data transfer requires two services, the polling system describing this FWLAN has rerouting once with probability one, i.e. $\epsilon_1 = 1$ and $N=2$.

{\it Notations:} In \richardCol{sections describing the FWLAN}, sample points in $\Delta$ are represented by $x$, $y$ and random points by $X$, $Y$. Sample and random points in the cyclic path ${\cal C}$ are represented by $q$ and ${Q}$ respectively. Subscripts $u$ and $d$ denote the uplink and downlink respectively.

{\it Assignment:} To point $q$ \richardCol{on} the ferry path is assigned a set of  points $I(q) \subset \Delta$. \mynCut{$I(q)$ is the set of points of $\Delta$ such that $q \in {\cal C}$ is the closest point   to them.} The ferry stops at $q$ if there is a user in $I(q)$ with either a downlink or uplink request. 
  This assignment is again a design issue, but we consider the nearest distance criterion. Namely if  $a$ is a corner point, then $I(a)$ is a rectangle otherwise it is a segment (see Figure \ref{Fig_RectFWLAN}). When the distance between two wireless nodes is small the \richardCol{signal attenuation is dominated by the distance-dependent propagation loss which justifies the above choice of assignment}.


{\it Arrivals:} Each data transfer consists of a fixed number of bytes $\eta$. The requests arrive according to a Poisson process and are associated with two marks: $X$ the position of the source  distributed as $P_X$ and $Y$  the position of the destination distributed as $P_Y$. We assume that $P_X$ is independent of $P_Y$.
 We assume uniform arrivals i.e.  $P_X \sim {\cal U} (\Delta)$ and $P_Y \sim {\cal U} (\Delta)$. This assumption is made to simplify the analysis and it is noted that Theorem 1 allows to analyze the system when $P_X$ and $P_Y$ are arbitrary distributions. 
 Every arrival in the  interval $I(q)$ is associated with an arrival at point $q$ of ${\cal C}$ in the equivalent polling system (see \cite{JournPaper} for more modeling details).  Thus,
 $$ P_{Q_0} (A) = P_X( \cup_{q \in A} I(q) )  \mbox{ for any Borel set } A \subset {\cal C},$$ gives the external arrival distribution. Similarly $P_{Q_1}(A) := P_Y( \cup_{q \in A} I(q) ),  $
for any Borel set $A \subset {\cal C}.$ Notice that the two arrival probabilities  are equal, i.e.  $P_{Q_0} = P_{Q_1} $. Denote the common measure by $\Psi := P_{Q_0}=P_{Q_1}$, which can easily be computed: 
\begin{eqnarray*}
\Psi (dq) & =&  \sum_{i=0}^3 p_0 1_{q_{i}} (q)   +  p_1 f_\psi (q) dq  \ \ \mbox{  where } \ \ p_0 := d^2  p_1, \ \ p_1 := \frac{1}{4 D_1D_2}, \\
q_0  &=& 0, \ \ q_1  =  2(D_1-d),  \ \ q_2 = q_1 + 2(D_2-d),  \ \  \\
 q_3 &=& q_2 + 2(D_1-d), \ \  \ \ \ \  \hspace{1.5mm} q_4 =  q_3  + 2(D_2-d) \ \ \ \ \mbox{ and }\\
 f_\psi (q) &=& \left \{ \begin{array} {llllll}
 d + (q  -q_i)                          &\mbox{if}&\hspace{-5mm}               q_i  <  q  <  q_i + D_1 - d    &\mbox{for some } 0 \le i \le 3 \\
 d + (q_i - q)                        &\mbox{if}&  \hspace{-5mm}             q_i - ( D_1 - d ) <  q  < q_i   &\mbox{for some } 1\le i\le 4 \\
   D_1                                         & \mbox{otherwise.} \hspace{-3mm}
\end{array} \right . 
\end{eqnarray*}
The arrival position measure $\Psi$ \Col{\Col{ has both a discrete and a continuous component,}}   even when $P_X$ and $P_Y$
   are continuous (uniform) distributions.  
  {\it Hence this example could not be modeled using any of the previous results.} \richardCol{However,} Theorem 1 \richardCol{enables} to model this example.

{\it Service times:} The ferry uses a wireless link to serve the users. We denote by $\kappa$ the decreasing function \richardCol{which maps the distance between a user and the ferry into the data rate at which they can communicate}. 

In particular, we assume that the \richardCol{data rate} depends only on the distance. \mynCol{ That is, we consider only the direct path and the associated path loss.} 
This assumption is valid because of the small distance between the two wireless nodes. Furthermore, assuming a height difference of $1$ between the receive and transmit \richardCol{antennas} and a path-loss factor $\beta$ (with $d$ the ground distance between \richardCol{the} user and the ferry and $\vartheta_{_P}$ the transmit power):
 \[
 {\kappa}(d) = \vartheta_{_P} (1 + d^2)^{-\beta/2} \mbox{ for every } d\ge 0.
 \]

\richardCol{Define} $q(x) : = \arg \min_{q \in {\cal C}_d} ||x-q||$ the closest point on the path, for any $x \in \Delta$.
A user at $x$ \Col{can communicate with \Col{the} ferry at data rate $\kappa (||q(x) - x||)$ and this is possible} when the ferry reaches $q(x)$. The time required for a data transfer, i.e. the service time equals:
\begin{eqnarray}
\label{Eqn_OverallServiceTim}
B(x) = \frac{\eta}{\kappa ( || q(x) -x|| )}  .
\end{eqnarray}
It is noted that the rate function $\kappa$ is bounded below as well as above in any bounded (rectangular) area and hence from (\ref{Eqn_OverallServiceTim}) the service time is bounded with probability one \Col{and so are its moments.}  \mynCut{With sufficient transmit power $\vartheta_{_P}$, the moments of the service times will be upper bounded by a constant that satisfies the conditions of Theorem 1.} \mynCol{When \richardCol{the} transmit power $\vartheta_{_P}$ is sufficiently high, the moments of the service times will be 
 upper bounded by a constant that satisfies the conditions of Theorem 1 and we are analyzing it under these conditions.  However we include the constant 
 ${\vartheta_P}$ into \richardCol{the} constant $\eta$ so that the notations are simpler.}

Every request first receives the uplink service at $X$ and the uplink service time equals $B(X)$. The downlink service \richardCol{time} is $B(Y)$. The two service times are independent of each other as required by Theorem 1.
The moments of the uplink or downlink service times depend upon the point $q \in {\cal C}$. For the uplink service,
\begin{eqnarray*}
    b_{0} (q) = b_u(q) &=& E[B(X) |  q(X) = q ] = E_X\left [ \left . \frac{\eta}{\kappa(  |q-X|  )} \right |  X  \in I( q)\right ] , \\
    b_{0}^{(2)}(q) = b_u^{(2)} (q)  &=&  E_X\left [ \left . \frac{\eta^2}{\kappa(  |q-X|  )^2} \right |  X  \in I(q) \right ].
\end{eqnarray*} \richardCol{The} downlink service moments $b_{1} = b_d$, $b_{1}^{(2)} = b_d^{(2)}$ can be defined in a similar way.
One can easily deduce that these service moments are the same for both uplink and downlink service. The common moments can be calculated as
(with $b := b_{1}  = b_{0}$ and  $b^{(2)} := b_{0}^{(2)} = b_{1}^{(2)}$): 
\begin{eqnarray}
\label{Eqn_FWLANbq}
b (q_i)   &=& \frac{\eta}{d^2} \int_0^{d} \int_0^{d}   (1+ x_1^2+x_2^2 )^{\frac{\beta}{2}} dx_1 dx_2  \mbox { for every } 0\le i\le 3 \\
b (q) &=&    \frac{\eta}{f_\psi(q)} \int_{-d}^{f_\psi(q)-d}  (1 + l^2)^{\frac{\beta}{2}}   dl   \hspace{3mm}     \mbox{ for every }  q  \notin \{q_i\} \mbox{ and }\\
b^{(2)} (q_i)  &=& \frac{\eta^2}{d^2} \int_0^{d} \int_0^{d}   (1+ x_1^2+x_2^2 )^{\beta} dx_1 dx_2  \mbox { for every } 0\le i\le 3 \\
b^{(2)} (q) &=&   \frac{\eta^2}{f_\psi(q)} \int_{-d}^{f_\psi(q)-d}  (1 + l^2)^{{\beta}}   dl   \hspace{3mm}     \mbox{ for every }  q  \notin \{q_i\} . 
\label{Eqn_FWLANbqEnd1}
\end{eqnarray}  
The average moments can  be computed using $\Psi$ and $\{ b(q) \}$ and $\{ b^{(2)} (q) \}$.  For example, because of symmetry,  
by interchanging the integrals and using a change of variables, the first moment simplifies to  (as $|{\cal C}| = 4(D_2-d) + 4 (D_1-d)$)
{\small
\begin{eqnarray}
{\bar b}  &=&  4 p_0 b(q_0) + 8 \eta p_1 \int_0^{D_1-d} \int_{-d}^{q} (1+l^2)^{\frac{\beta}{2}} dl  dq
+  2 \eta p_1  2(D_2-D_1)  \int_{-d}^{D_1-d} \hspace{-3mm}(1 + l^2) ^{\frac{\beta}{2}} dl  \nonumber \\
&=& \frac{\eta}{D_1D_2} \left ( |{\cal C}| h_{\beta, d} +8 \frac{(1+(D_1-d)^2)^{\frac{\beta}{2}+1} - (1+d^2)^{\frac{\beta}{2}+1}}{\beta/2+1} 
 + 4 g_{\beta,d} \right ) \mbox{ with}  \\
h_{\beta,d} &:= &
 \int_0^{D_1}   (1+ (l-d)^2 )^{\frac{\beta}{2}} dl  \ \ \mbox{ and } \ \
g_{\beta,d} \ \ :=  \ \
 \int_0^{d} \int_0^{d}   (1+ x_1^2+x_2^2 )^{\frac{\beta}{2}} dx_1 dx_2. \nonumber
 \label{Eqn_FWLANbqEnd}
\end{eqnarray} }
{\it 
Hence the FWLAN can be modeled by a \cts polling system with rerouting with parameters as computed above.
Theorem 1 can be applied and the stationary expected virtual workload of the FWLAN can be calculated using
(\ref{Eqn_Vmix}) for any given cyclic path ${\cal C}_{d}$ and the corresponding line segments $\{ I_q \}_{q \in {\cal C}}$.  
}
Our goal is to find the optimal rectangular ferry path: 
\[
d^*_V = \arg \min_{0 <  d \le D_1} V_{fwlan} (d)
\]  where $V_{fwlan} (d)$ represents the expected stationary workload of the FWLAN when the ferry moves in ${\cal C}_d$.  \richardCol{$V_{fwlan}(d)$} is calculated by substituting the common moments $\{ b(q)\}$, $\{ b^{(2)} (q) \}$ and $P_{Q_0} = P_{Q_1} = \Psi$ into the
stationary expected workload given by (\ref{Eqn_Vmix}), which simplifies to ($\rho = 2\lambda {\bar b}$):
\begin{eqnarray}
\label{Eqn_Vfwlan}
V_{fwlan} (d)
&=& \frac{\rho \lambda}{1-\rho} \left ( {\bar b}^{(2)} +  
{\bar b}^2 \right )
+   \rho  |{\cal C}|  \alpha^{-1}    \frac{2+\rho}{4(1-\rho)}
 \\
%
&& 
+  { \lambda \alpha^{-1}}
\left ( E_\Psi [ Q b({Q})]  - {\bar b} E_\Psi [Q]  + |{\cal C}| E_\Psi [{\hat b} (Q) ] \right ).
\nonumber 
\end{eqnarray}  Here $E_\Psi$ represents the expectation  w.r.t. $\Psi$. For any integrable function $g$,
$$E_\Psi[g(Q)] = \int_0^{|{\cal C}|} g(q) \Psi (dq)  = \sum_{i=0}^3 g(q_i) p_0 + p_1 \int_0^{|{\cal C}|} g(q) f_\psi(q) dq.$$
We obtain the stationary expected workload performance of the FWLAN for every value of $d$. This expression cannot be simplified and one has to compute the optimal path \richardCol{numerically}.

\subsubsection*{Numerical Examples} 
\begin{figure}[!htbp]
\centering
\includegraphics[width=1\columnwidth]{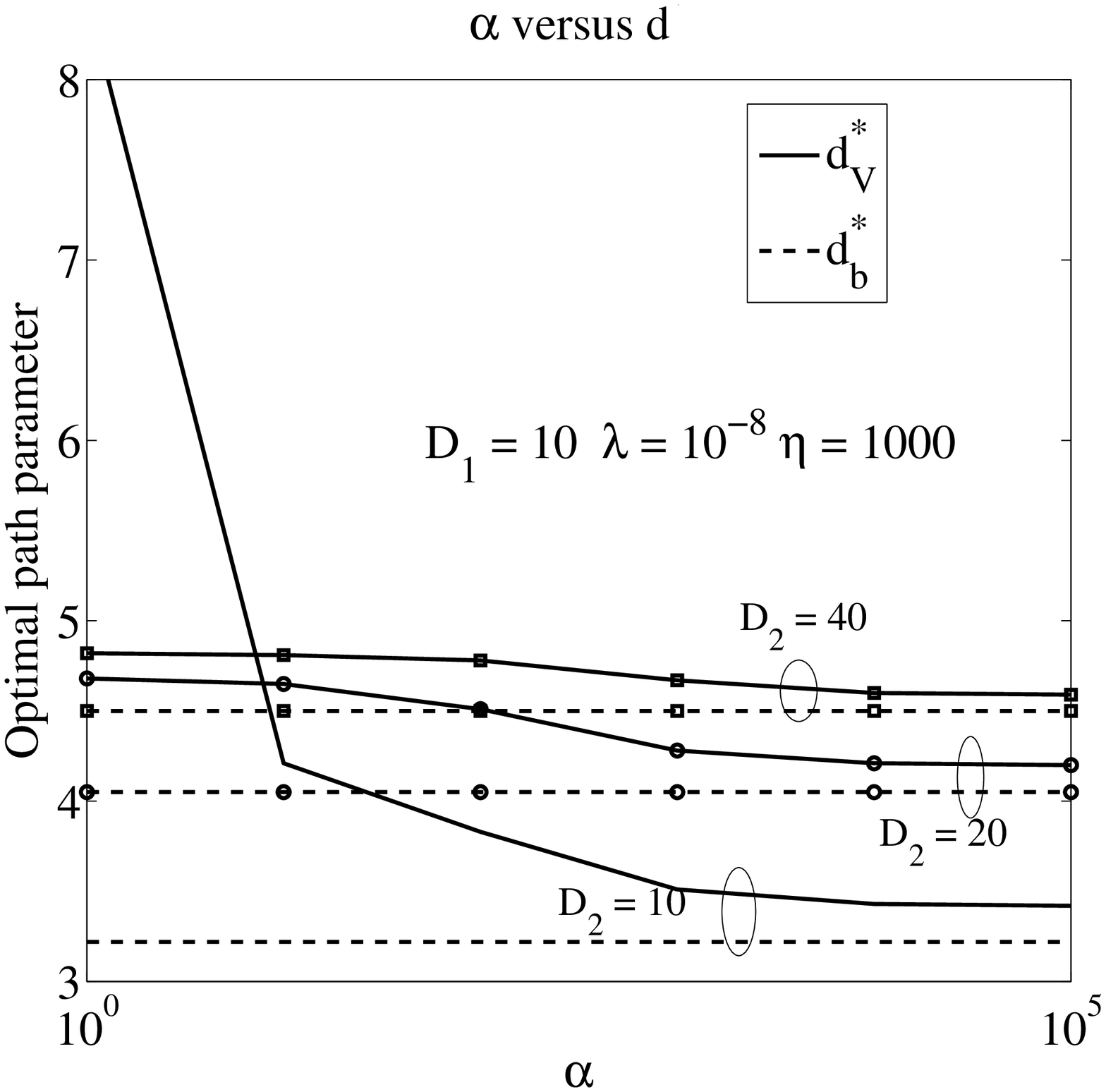}
\caption{Optimal path for a Ferry  moving  in a rectangular area \label{Fig_Rectexample1}. Minimizer for workload: $d_V^* := \arg \min_{0\le d \le D_1} V_{fwlan} (d)$, minimizer for the first service moment: $d_b^* := \arg \min_{0\le d \le D_1} {\bar b}$.}
\end{figure}

We \richardCol{derive} the optimal path parameter $d_V^*$, the minimizer of the expected workload $V_{fwlan}$, for a few examples using numerical methods. 
We also compute $d_b^*$ the \richardCol{minimizer} of the first moment of the service time, ${\bar b}$ (see Figure \ref{Fig_Rectexample1}). 
In Figure \ref{Fig_Rectexample1} we plot the optimal parameter as a function of the ferry speed $\alpha$ for different values of the longer dimension $D_2$. 
We set $\beta = 2$, $\eta = 1000$, $D_1 = 10$ and $\lambda = 10^{-8}.$ 	 We notice that the two optimizers $d^*_V$ and $d^*_b$ \richardCol{become} closer to each other when the speed $\alpha$ or the longer dimension $D_2$ increase. This can be explained by observing equation (\ref{Eqn_Vfwlan}). The first term of   (\ref{Eqn_Vfwlan}) depends upon $d$ mainly via the service moments and \richardCol{is minimized} by a $d^*$ close to $d_b^*$. The \richardCol{other} terms are proportional to $\alpha^{-1}$ and \richardCol{they vanish} for large ferry speeds. Another important observation is that the optimal $d^*$ approaches $D_1/2$
as $D_2$ increases.

\begin{figure}[!htbp]
\centering
\includegraphics[width=1\columnwidth]{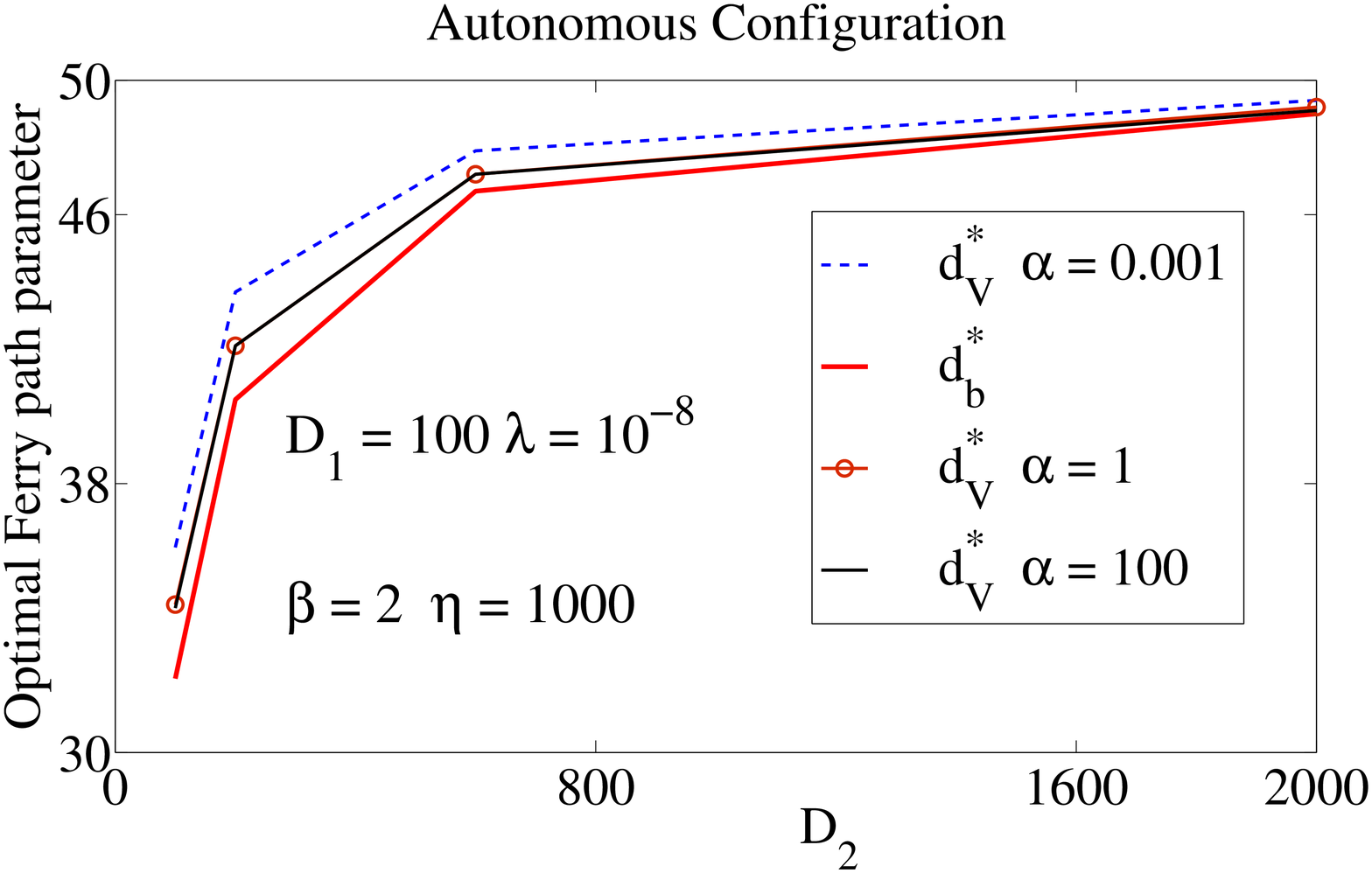}
\caption{  Optimal path  parameter for the autonomous architecture  \label{Fig_Rectexample2}}
\end{figure}

	Our conclusions are reinforced by Figure \ref{Fig_Rectexample2}. Here we plot the optimal path parameter as a function of $D_2$ for various ferry speeds $\alpha$.  In this example we set $D_1 = 100.$ We notice that the optimal path parameter converges to $D_1/2$ as the larger dimension increases. 
\richardCol{ Furthermore the minimizers are close to $d_b^*$ as either the speed $\alpha$ or the dimension $D_2$ increase.} We can make two interesting observations:

O.1) The optimizer $d^*_V$, in most of the cases, is  close to  the optimizer of the first moment of the overall service times ${\bar b}$;   

O.2) { For thin or long strips of area  the optimal ferry path parameter is one fourth  of the lower dimension.  }

\begin{figure}[!htbp]
\centering
\includegraphics[width=1\columnwidth]{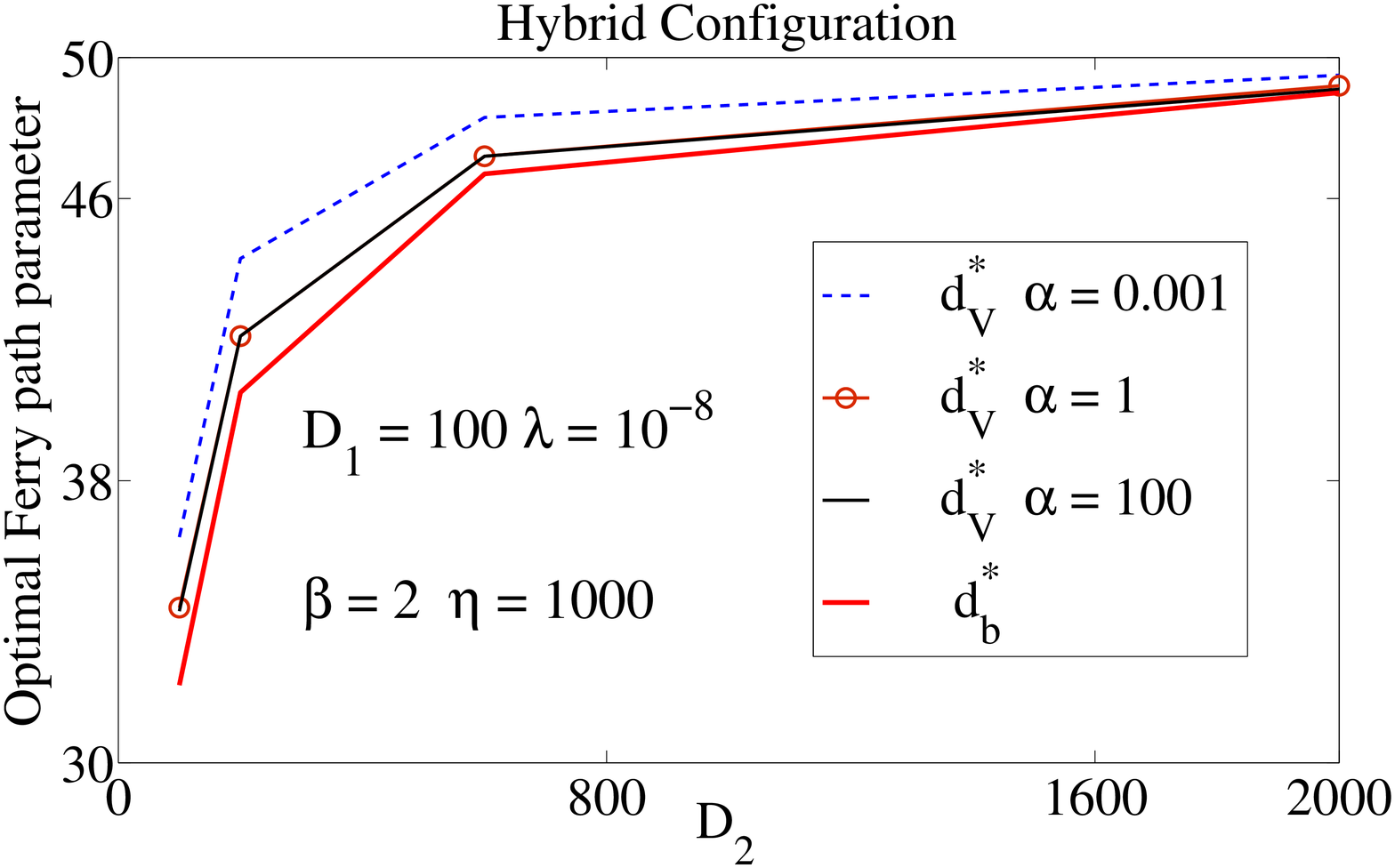}
\caption{Optimal path  parameter for the hybrid architecture  \label{Fig_Rectexample3}}
\end{figure}

\subsection{Hybrid Architecture}
The previous architecture of FWLAN is called the ``autonomous architecture'' as the ferry itself manages the local communications. In certain situations, 
the ferries are built with minimal intelligence, and the rerouting task is done by the base station (BS). This architecture with a finite number of ferry stops is discussed in \cite{itc}. In this case every local data transfer  happens in three phases. First there is an uplink service from the source node to the ferry. Then the ferry contacts the BS to obtain the address of the sink node. Finally the ferry travels to the sink location and there is a downlink service to the sink node. All the other details remain as in the previous section. 

\richardCol{Most of the mapping details between the system and the corresponding polling system remain the same as in the previous section.} We point out only the differences. Every arrival \richardCol{requires} $3$ services: uplink, downlink and BS-transfer service.
Thus there are $2$ reroutings, i.e. $N=3$, with $\epsilon_1 = \epsilon_2 = 1$. 
{\it Note here that \mynCut{our} previous results (\cite{Value}) could not model this architecture as $N=3$.}
However, the polling systems considered in this paper can model this architecture.
As in the previous section, $P_{Q_0}$ and $P_{Q_2} $ are defined using $P_X$ and $P_Y$  respectively. $P_{Q_1} $ is concentrated only 
at the BS (which we assume is associated with the point $0$ of the ferry path), i.e. $P_{Q_1} (\{0\}) = 1$.  \mynCol{This is another reason why the previous systems cannot model this FWLAN.} \Cut{The service first moments $b_{0} = b_{2}$ are the same as the common moments $b$ defined in the two previous sections and so are the second moments.}
\Col{The moments of the service times \Col{for} the first and the last service are equal and they 
equal the common moments given by equations (\ref{Eqn_FWLANbq})-(\ref{Eqn_FWLANbqEnd}) in the previous section, i.e. $b_0(q)=b_2(q) = b(q)$, ${\bar b}_0 = {\bar b}_2  = {\bar b}$
and $b^{(2)}_0 (q) = b^{(2)}_2 (q)$ for all $q$.}
 If the BS is at \Col{a} distance $d_{bs}$ from the ferry path,  then \Col{the  moment corresponding to \Col{the} intermediate service or the \Col{second} service equals}
$$
 b_{1} (0) = b_{bs} := \frac{\eta_1}{\kappa(d_{bs})}, \ \   b_{1}^{(2)} (0) = b_{1} (0)^2 \mbox{ and } {\bar b}_1 = E_{Q_1}[b_1 (Q_1)] = b_1 (0) ,$$ where $\eta_1$ can be different 
from $\eta$.  
In this case, 
$$
\rho = 2\lambda {\bar b} + b_{bs} \lambda,  \mbox{ and }
  {\hat \rho } (q) = b_{bs}  \lambda + 2 \lambda {\hat b} (q) \mbox{ for all } q. 
$$
The stationary workload for the FWLAN with hybrid architecture is given by: 
\begin{eqnarray*}
V_{fwlan}^{hybrid} \hspace{-2mm}  &=& \frac{\rho \lambda  \left (2 {\bar b}^{(2)} + 2  {\bar b}^2 + b_{bs}^2 + 4 b_{bs} {\bar b}\right )  }{2 (1-\rho)}
+ \frac{  |{\cal C}| \alpha^{-1} \left (2 \rho + 2 \lambda^2 {\bar b}^2 -  \lambda^2  b_{bs}^2 \right ) }{4 (1-\rho)}   \\
&& +  \frac{ \lambda |{\cal C}| \alpha^{-1} (b_{bs}+{\bar b}) }{1-\rho} 
+ \lambda \alpha^{-1} \left ( E_\Psi [Q b(Q)] - ({\bar b} + b_{bs} ) E_\Psi [Q] \right ) \\ && - 
 \frac{2 \lambda^2 |{\cal C}| \alpha^{-1} (b_{bs}+{\bar b})  E_\Psi [{\hat b}(Q)] }{1-\rho} .
\end{eqnarray*}
When $b_{bs} = 0$, i.e. \richardCol{the} time needed for the BS to transfer the address of the sink node to the ferry is negligible,
\begin{eqnarray}
\label{Eqn_Vfwlanhyb}
V_{fwlan}^{hybrid}   &=& \frac{\rho \lambda  \left ( {\bar b}^{(2)} +  {\bar b}^2 \right )  }{ (1-\rho)}
+ \frac{  |{\cal C}| \alpha^{-1} \left ( 2 \rho +  \lambda^2 {\bar b}^2 \right ) }{2 (1-\rho)}   \nonumber \\
&&  
+ \lambda \alpha^{-1} \left ( E_\Psi [Q b(Q)] - {\bar b}  E_\Psi [Q]   -  \frac{\rho}{1-\rho}  |{\cal C}| E_\Psi [{\hat b}(Q)]  \right ) .
\end{eqnarray}
\richardCol{This happens when the BS is on the ferry path.} 


\subsubsection*{Numerical Examples}

We \richardCol{still consider} the example of Figure \ref{Fig_Rectexample2} for the hybrid architecture in  Figure \ref{Fig_Rectexample3}. We notice that the minimizers behave \richardCol{as in} the case of autonomous architecture. In fact, they are very close to the minimizers for the autonomous architecture
 when the ferry speed or $D_2$ are large. We notice that the two optimizers differ only when the ferry speeds are moderate and $D_2$ is close to $D_1$. The difference is not large though. This is in line with the two expressions for the workload (\ref{Eqn_Vfwlan}) and (\ref{Eqn_Vfwlanhyb}). The two workloads are 
close to each other for large values of ferry speeds.  
  \richardCol{We make a third observation:}
 
 O.3)  The optimal path remains more or less the same irrespective of whether the ferry manages the local communications completely on its own or with the help of the BS.



\section{\Col{Application:}  Distributed waste collector}
\label{section_collect}

Consider a large area in which waste is generated continuously at various locations. A dedicated vehicle moves in a closed path collecting the waste continually and discards it at a given collection point. We would like to find the optimal collection point, given the waste generation rate and distribution. 

This scenario can occur for example in a production unit, where production of intermediate products is spread over a large area. The waste generated from the production of intermediate products needs to be cleared out. Another example would be an area hit by a natural disaster. People in charge of the rescue operation must clear out the debris. The debris are collected in packets and the packets are placed in the path of the collecting vehicle which picks it up. The vehicle picks up the packets when it encounters them on its path. The number of pickup points need not be finite and pickup can occur anywhere on the path of the vehicle. Hence we need \cts polling systems to model this situation. Another example is a sensor network in which sensors are distributed over a large area. \richardCol{The sensors gather information and send it to a center which aggregates it}. In FWLANs as well we might want to find the optimal location of the BS, given the distribution of the arrivals. For example, on a campus\richardCol{,} a particular department might require more data transfers than others. In this case the arrival distribution \richardCol{is not uniform} and one needs to design an optimal \richardCol{collection} point given this distribution. 

Consider \richardCol{a} vehicle moving at unit speed on a cyclic path of unit length collecting goods from users along the path. Every pickup needs a fixed amount of time $p$. The goods \richardCol{must} be discarded at one fixed point $x_d$.
Discarding an item requires a fixed amount of time $d$. The requests arrive according to a Poisson process with arrival rate $\lambda$ on the cyclic path according to some distribution $f_Q (x) dx$. Given the arrival density $f_Q$, we would like to \richardCol{derive} the optimal \richardCol{collection} point $x_d$ to minimize the total expected workload in the system.

\subsection{Mapping to a \cts polling system}
We once again map this waste collection system to an appropriate polling system and obtain its performance using Theorem 1. 
    The vehicle collecting the waste represents the server of the polling system. The new/external arrival distribution is given by $f_Q (x) dx$, i.e. in the notations of the \cts polling system, $P_{Q_0} (dx) = f_Q (x) dx.$ Every arrival gets rerouted to the location where the waste is discarded. The job is over when the waste has been discarded.
    Namely after the first job (collecting the waste) all the requests are rerouted to a single point $x_d$ and hence the arrival distribution $P_{Q_1}$ is not a continuous distribution. \richardCol{As} in the case of the previous example, this situation cannot be modeled by 
any of the previously studied polling systems  (e.g., \cite{JournPaper,Value,Coffman1,Coffman2,Eliazar,Eliazar1,Kroese1} etc). However this paper supports mixed arrival distributions as well as rerouting \richardCol{and} this example can be analyzed. 
In this case, $$\epsilon_1 = 1, \ \  N = 2, \ \ P_{Q_0} (dx) = f_Q (x) dx \mbox{ and } P_{Q_1} (\{ x_d \}) = 1 . $$ 
    
Every pickup \richardCol{takes} a fixed amount of time $p$ and discarding an item \richardCol{takes} a fixed amount of time $d$. Hence 
$$
{\bar b}_{0} (q) = p, \ \  {\bar b}_{0}^{(2)} = p^2,  \ \ {\bar b}_{1} (q) = d \ \ {\bar b}_{1}^{(2)} = d^2 \mbox{ for all } q. 
$$  
\subsection{Stationary expected virtual workload and optimization}
Theorem 1 can be applied to the distributed waste collector and using (\ref{Eqn_MainFormula}) its stationary workload equals:
\begin{eqnarray}
 V_{WC} 
&=&  \frac{\rho \lambda (p^2 + d^2)}{2(1-\rho)}   +  \frac{\rho \lambda p d}{1-\rho}  
+   \frac{\rho       } {2 }
+   \frac{\lambda  d}{1-\rho}  \left ( x_d - E_{Q_0} [Q] + (1- F_Q (x_d) ) \right ) 
\nonumber \\
&& \hspace{-16mm}
+\frac{\rho  }{1 - \rho }  \int_0^{1} \int_0^{1} 
\lambda \left (  p \left ( F_Q (y) -  F_Q (q) + 1_{\{ y < q \}} \right ) + d \left ( 1_{\{ x_d \le y \}} - 1_{\{ x_d \le q \}}
+   1_{\{ y < q \} }   \right ) \right )  \nonumber \\ && \hspace{100mm} f_Q(q) dq     dy   \nonumber 
\end{eqnarray} 
with load factor
$
\rho = \lambda (p + d)
$ and with $F_Q (q) := P_{Q_0} ([0,q])$ representing the cumulative distribution function of $f_Q$. We would like to choose the collection point $x_d$ \richardCol{minimizing the expected} workload $V_{WC}$. After simplifying $V_{WC}$ and considering only the terms \richardCol{which depend on} $x_d$,  \myCol{ we obtain:}
\begin{lemma}
Given the cumulative distribution function $F_Q (.)$ of the waste generation positions, the location optimizing the stationary workload equals:
\begin{eqnarray}
\label{Eqn_WCOpt}
x_d^* := \arg \min_{ 0 \le x_d \le 1} V_{WC} = \arg \min_{ 0 \le x_d \le 1}  \left[ x_d - F_Q (x_d) \right ].  \hspace{20mm} \Box
\end{eqnarray}
\end{lemma}

By the above lemma, the \richardCol{collection} point can be anywhere on the path if waste is generated uniformly (in this case $F_Q (x) = x$ for all $x$). \richardCol{By the usual first-order derivative condition,} the optimal point to discard the waste satisfies:
$$
f_Q (x_d^*) = 1 \ \  \  \ \mbox{ if } \  \ \frac{df_Q}{dx} (x_d^*) > 0.
$$


\section{Proof of Theorem 1} 
\label{section_proof}

\richardCol{We obtain the proof in three steps using a discretization approach as in \cite{JournPaper}. There are large differences in the proof with respect to~\cite{JournPaper} because of the presence of atoms in the arrival position measure. The existence of atoms }
(see (\ref{Eqn_Arrivalprobs})) allows the possibility of two or more customers waiting at the same point (at one of  $\{q_{i,j}\}$ of (\ref{Eqn_Arrivalprobs})) \mynCol{and forming waiting lines}. This possibility complicates the analysis to a great extent and requires major changes in the steps 2 and 3 of the proof.  In \cite{JournPaper} we obtain the convergence of certain second moments in supremum/uniform norm (with respect to all the points on the circle) in an intermediate step. However this strong form of convergence is not possible here because of the existence of atoms. 
We circumvent this problem by obtaining a \Col{(somewhat}) weaker form of convergence ($L^1$ convergence) \richardCol{which still proves the announced result.}  
 
\noindent \underline{Step 1)Discretization:}  The \cts polling system with rerouting
 is converted to an appropriate discrete polling system with rerouting in subsection \ref{section_discrete}. For the discrete system, the pseudo conservation laws apply \richardCol{and} the stationary expected virtual workload is known (see \cite{SLF}).

For $q \in {\cal C}$ let $\delta^\sigma(q)$ denote the point at which users  of $q$ will be served in the discretized system with $\sigma$ discretization levels. Let $\delta^\infty$ represent the $\delta$ \richardCol{function} for the \cts system. Note that $\delta^\infty (q) = q$ for all $q$ so that $\delta^\infty$ is the identity map.

\noindent\underline{Step 2)Fixed point equations:}\hspace{2mm}
\richardCol{We are interested in the stationary moments of the cycle time with respect to $q$, which is the time period between two successive visits at location $\delta^\sigma(q)$.} We express those moments as fixed points (in the space of left continuous and right limit functions) of an \richardCol{affine} operator in subsection~\ref{section_fixedpoint}. We define two common  operators $({\cal F}, {\Theta})$ in subsection  \ref{section_fixedpoint}, 
both of which are parametrized by $\sigma$.   We study the continuity of $({\cal F}, {\Theta})$ with respect to $\sigma$ and show that the stationary moments of the discrete system converge to \richardCol{the stationary moments of} the \cts system as $\sigma \to \infty$.

\noindent\underline {Step 3)Alternate expression for the virtual workload:} \hspace{1mm}
 We express the stationary expected virtual workload  
     in  terms of the  stationary moments of Step 2. It should be noted that this expression cannot be computed easily
and is used only for the purpose of the proof. We show that the stationary expected virtual workload of the discrete system converges to the stationary expected virtual workload of the \cts system in subsection \ref{section_workload}.

\subsection{Discretization}
\label{section_discrete}
For each integer $\sigma$, we construct  a discrete polling system with  $N\sigma$  queues such that the limit (as $\sigma \to \infty$) of the performance measure of this system \Col{equals} \richardCol{the performance measure} of the \cts system. The number of queues \richardCol{for serving} the external arrivals \Cut{to  the system} is $\sigma$ while the rerouted users are served in the remaining  $(N-1)\sigma$ queues.  \richardCol{There are $\sigma$ queues to serve the $j$-rerouted users (i.e. the users that already received $j$ services and are awaiting the $(j+1)$-th service), for each $1 \le j < N$.}
The circumference $|{\cal C}|$  is divided  into $\sigma$ \richardCol{segments of equal length} $\{ I_i\}_{i=1}^{\sigma}$ with $I_1 = [0, {\cal C}/\sigma]$.
External arrivals are as in the \cts system.
Users arriving in area $I_i$ are treated as though arriving in queue numbered ${N(i-1) +N-j}$ (for ``$j$-rerouted'' users)
or ${Ni}$ (for  external users).
For every $i$,  the server stops upon reaching the starting point, $i^\sigma := (i-1)|{\cal C}|/\sigma$, of $I_i$ and serves the users  of  $I_i$
  before moving further.
Hence, $\delta^{\sigma} (q)$, the point at which the server serves the users who arrived at $q$, equals
\begin{eqnarray}
\label{Eqn_deltasigma}
\delta^{\sigma} (q) = \left \{ \begin{array}{lllll}  \sum_{i=1}^{\sigma} i^\sigma   1_{\{ q \in I_i \}} \mbox{  with }
I_i  := \left [ i^\sigma,   (i+1)^\sigma  \right )  & \mbox{ when } &\sigma < \infty  \\
q     & \mbox{ when } &\sigma = \infty.
\end{array}
\right .
\end{eqnarray}
The server first \Col{serves} the  ``$(N-1)$-rerouted''  queue
($N(i-1) +1$ \Col{numbered} queue), using gated service:
  the server \Col{serves} all the users that were rerouted for the $(N-1)$-th time before
the server reached $i^\sigma$.  Then it serves the $(N-2)$-rerouted users and so on until it \Col{\Col{serves}} the $1$-rerouted queue, using gated service.
After all the rerouted queues, the server \richardCol{serves} the ``external'' queue once again using gated service. Namely it serves all the external arrivals of $I_i$ that arrived \Col{before completing the service of} all the rerouted users of $I_i$. 

Within a queue, the server \Col{serves} the users using a special order which we call  {\it arrival position order}. In this special order, the users within
a queue are served in the order of their distance from the stop $i^\sigma$ of the server, i.e.  {\it the user at  minimum distance is served first.} Furthermore  the users waiting at the same point are served in {\it FIFO} order. The users are served in almost the same order as in a \cts system. The main difference
between the \cts system and the discretized system is that some of the users are postponed to the next cycle in the discretized system. This is because of the combination of discretization and gated service. We will \richardCol{prove} that the effect of this difference reduces to zero as $\sigma$ tends
to $\infty$ \mynCol{(while proving Theorem 1).}

\newcommand{\qnum}[2]{ \ensuremath{{ {#1_{_{_{\hspace{-0.5mm}(#2)}}}}  }} }

Define $\qnum{i}{j} := N(i-1)+N-j$, \richardCol{the index of the queue for $j$-rerouted users standing at $i |{\cal C}|/\sigma$.} \richardCol{After their first service, the external arrivals (at one of the $\qnum{i}{0}$  numbered queues) are either rerouted to one of the $\qnum{k}{1}$ queues (with $k \le \sigma$) or leave the system}. Similarly users of queue $\qnum{i}{j-1}$ are either rerouted to queues  \RevCol{$\{\qnum{k}{j}\}_k$} or exit the system. These transitions occur according to the following probabilities (by independence):
\begin{eqnarray*}
P_{ \qnum{i}{j-1}, \qnum{k}{j} }
&=& Prob \left(\mbox{User in $[i^\sigma, (i+1)^\sigma]$  } \mbox{  $j$-rerouted to $[k^\sigma, (k+1)^\sigma]$} \right) \\
&=& \epsilon_j P_{Q_j} (I_k)  \mbox{ for all }  1 \le j \le N-1, \mbox{  } i, k \mbox{ and}
\\
P_{\qnum{i}{j}, \qnum{k}{j'}}  &=&
P_{\qnum{i}{j}, \qnum{k}{0}} =
 0  \mbox{ for all } i,k, j \mbox{ and } j' \mbox{ with } j' \ne j+1.
\end{eqnarray*}
 Poisson arrivals with intensity $\lambda$ \richardCol{enter one of} the $\qnum{i}{0}$ queues. The arrivals
 in  $I_i$ form the external arrivals to queue  $\qnum{i}{0}$. The rate of external arrivals at the different queues are:
$$ \lambda_{\qnum{i}{0}} = \lambda P_{Q_0} (I_i) \mbox{ and }   \lambda_{\qnum{i}{j}} =0  \mbox{ for all } i \le \sigma \mbox{ and } 0 < j < N.  $$
A user at queue $\qnum{i}{j}$ \richardCol{has a service time} $B_{Q_j}$ and this service time is conditioned on the event that the arrival is in $I_i$.
Thus, the service time moments at the different queues are:
\begin{eqnarray*}
\begin{array}{lllllll}
b_{\qnum{i}{j}}  &=& \frac{E [ B_{Q_j} 1_{\{ Q_j \in I_i \} }]}{P_{Q_j}(I_i)}    & \hspace{5mm} &
b^{(2)}_{\qnum{i}{j}}  &=& \frac{ E [ B_{Q_j}^2 1_{\{ Q_j \in I_i \} }]}{P_{Q_j}(I_i)}
 \mbox{ for all } 0 \le j < N.
\end{array}
\end{eqnarray*}
	Every external arrival can potentially cause several rerouted arrivals. Hence one needs to estimate 
the total rate at which arrivals occur,  including both external arrivals and reroutings. This computation is obtained using balancing arguments and is explained in \cite{SLF}. Similarly the workload incurred by an arrival is equal to the workload of the first service plus the workload of subsequent services. \richardCol{We also calculate the total service time requirements.} The overall arrival rates $\gamma_i$ (resulting after rerouting) can be calculated solving  \cite[eqn. (2.1)]{SLF} inductively as (we recall that ${\hat \epsilon}_j = \Pi_{k=0}^j \epsilon_k$):
\begin{eqnarray*}
\gamma_{\qnum{i}{0} }\hspace{-2mm} &=& \hspace{-2mm} \lambda_{\qnum{i}{0}} 
+ \sum_{\myCol{k=1}}^{N\sigma} \gamma_k  P_{k, \qnum{i}{0}} \implies \gamma_{\qnum{i}{0}}  = \lambda_{\qnum{i}{0}}   \mbox{ for all } i \mbox{ and so}\\
\gamma_{\qnum{i}{j}}\hspace{-2mm} &=& \hspace{-2mm} \lambda_{\qnum{i}{j}} + \sum_{k=1}^{N\sigma}  \gamma_k P_{k, \qnum{i}{j} }
= \epsilon_j  P_{Q_j}(I_i)  \sum_{l=1}^{\sigma}  \gamma_{\qnum{l}{j-1}}
 \implies \gamma_{\qnum{i}{j}} =   {\hat \epsilon}_j  \lambda P_{Q_j}(I_i).   
\end{eqnarray*}
The overall service time requirements resulting from the first and the possible \mynCut{second service} \mynCol{subsequent services},  ${\tilde b}$,
 can be calculated as below 
(solving \cite[equations 2.3 and 2.4]{SLF} by induction starting with $j=N-1$ and with 
$\check {\epsilon}_j^k := \Pi_{i=j}^{k} \epsilon_i$):
\begin{eqnarray*}
{\tilde b}_{\qnum{i}{j}} \hspace{-1mm}&=\hspace{-1mm}& b_{\qnum{i}{j}} +\sum_{k=j+1}^{N-1} {\check \epsilon}_{j+1}^{k} {\bar b}_{k}  \\ 
{\tilde b}^{(2)}_{\qnum{i}{j}} \hspace{-1mm}&=\hspace{-1mm}&{b}^{(2)}_{\qnum{i}{j}}   +  \hspace{-1mm}  \sum_{k=j+1}^{N-1} {\check \epsilon}_{j+1}^{k} {\bar b}^{(2)}_{k}   
+ 2 b_{\qnum{i}{j}} \hspace{-1mm}  \sum_{k=j+1}^{N-1} {\check \epsilon}_{j+1}^{k} {\bar b}_{k}
+ 2  \hspace{-1mm} \sum_{l=j+1}^{N-2} \sum_{k=l+1}^{N-1} {\check \epsilon}_{j+1}^{l} {\bar b}_{k} {\bar b}_{l}. 
%
\end{eqnarray*}
\begin{eqnarray*}
\mbox{Define, } \rho_{\qnum{i}{j}} &:=&  \gamma_{\qnum{i}{j}} b_{\qnum{i}{j}}  \mbox{ and }
\rho \ \ = \ \ \sum_{i,j}  \rho_{\qnum{i}{j}}  \ \
= \ \  \lambda {\bar b}_{0} + \sum_{j=1}^{N-1} {\hat \epsilon}_j \lambda {\bar b}_{j}. \hspace{5mm}  
\end{eqnarray*}
Note that $\rho$ \richardCol{does not depend on} $\sigma$. $\rho$ is the total workload in the system. The discrete system is stable only when $\rho < 1$ (\cite{SLF}). This condition is guaranteed because we study \cts polling systems \Col{under the assumption that they are stable}. So the stationary moments of Theorem \ref{Thrm_FirstMoments}  (given in  subsection \ref{section_fixedpoint}) exist and by the same theorem it is possible if and only if $\rho < 1$.

\noindent\underline{Workload of the discrete system:}
We have a stable polling system with $N \sigma$ queues. \richardCol{There are $\sigma$ 
queues experiencing gated service with external arrivals and $(N-1)\sigma$ 
queues also experiencing gated service but with only rerouted users.} \richardCol{Furthermore the walking times between queues are fixed.} \myCol{Let $r_k$ denote the walking time between queue $k$ and $k+1$. Between the queues of the
same stop the walking time is zero so $r_{\qnum{i}{j}}=0$ when $j > 0$ and $r_{\qnum{i}{0}} =  {|\cal C|}\alpha^{-1}
/\sigma$.}
\noindent This type of discrete polling system with rerouting is considered in \cite{SLF}.
  By the pseudo Conservation Laws  of \cite{SLF} the  expected stationary  workload of  the $\sigma$-polling system with rerouting
 is (from \cite[eqn. (6.4)]{SLF} after removing the zero terms):
\begin{eqnarray}
\label{Eqn_DiscreteCmbinedMain}
\label{Eqn_MixedService}
V_{rrt}^\sigma &=&    \frac{  \sum_{i=1}^{{\sigma}} \lambda_{\qnum{i}{0}} {\tilde b}_{\qnum{i}{0}}^{(2)} }{2 (1-\rho)}
- \sum_{j=0}^{N-1} \sum_{i=1}^\sigma \gamma_{\qnum{i}{j}} 
\left [ \frac{b_{\qnum{i}{j}}^{(2)}}{2} + \left  ({\tilde b}_{\qnum{i}{j}} - b_{\qnum{i}{j}} \right ) b_{\qnum{i}{j}}\right ] 
\nonumber
\\ &&
+   \frac{\rho  |{\cal C}|  \alpha^{-1}   } {2 }   
+   \frac{|{\cal C}|}{1-\rho} \hspace{-1mm} \sum_{i=1}^{\sigma} \frac{\alpha^{-1}}{\sigma} \sum_{l=1}^{\sigma} \lambda_{\qnum{l}{0}}   {\tilde b}_{\qnum{l}{0}}
\hspace{-1mm} \sum_{k = Nl}^{Ni} \rho_k
\nonumber \\ &&
+   \frac{1}{1-\rho} \sum_{j=0}^{N-2} \sum_{i=1}^{\sigma}  \sum_{l=1}^{\sigma} \gamma_{\qnum{i}{j}}  \epsilon_{j+1} P_{Q_{j+1}} (I_l) 
  {\tilde b}_{\qnum{l}{j+1}} \myCol{
\sum_{k = \qnum{i}{j}}^{\qnum{l}{j+1}-1}  r_k .}  
\end{eqnarray}
The results of  \cite{Boxma,Boxma1,Boxma2,SLF} are valid for any work conserving order at each queue and hence the results are also valid
for our {\it arrival position order.} 

\richardCol{We will prove that the limit of the expected stationary virtual workload in the discretized system, $V_{rrt}^\sigma$, equals the expected stationary virtual workload of the \cts system. We prove this fact to prove Theorem \ref{Thrm_MixedService} in the next two subsections.} 
 We conclude this subsection by computing the limit of (\ref{Eqn_DiscreteCmbinedMain}):
\begin{lemma}
\label{Lemma_LimVmixsigma}
The limit of $V_{rrt}^\sigma$ (\ref{Eqn_DiscreteCmbinedMain}) equals $V_{rrt}$ given by (\ref{Eqn_Vmix}) of Theorem 1.   
\end{lemma}
\myCol{{\bf Proof:}  The proof is in Appendix A. \hfill{$\Box$}}

\subsection{Fixed point equations}
\label{section_fixedpoint}
\richardCol{Let $0$ be {\it any arbitrary point of the circle, ${\cal C}.$ } } We call ``cycle time with respect to 0'' the time period between two successive visits of the server to the point $0$. Let $\phi_n^\sigma(q)$ denote the time at which the server starts the service of the external queue  ($\qnum{i}{0}$ numbered queue with only external arrivals) in the $n^{\mbox{th}}$ cycle, to which the point $q$ belongs. \richardCol{For the} \cts system, this corresponds to the time needed for the server to reach point $q$, complete the service of all the rerouted users (if any) at point $q$ and start serving the external users of $q$, in the $n^{\mbox{th}}$ cycle.
 Let $T_n^\sigma(q) :=  \phi_n^\sigma (q) - \phi_n^\sigma(0)$, \richardCol{denote} the time \richardCol{taken by the server to travel from $0$ to $q$ while serving all the users on the way up to the start of service at the external queue for point $q$, in the $n^{\mbox{th}}$ cycle.}
Let ${\cal T}_{Q_0} ([a,c], T)$ \richardCol{denote} the total workload of the external arrivals that arrived in $[a,c] \subset [0, |{\cal C}|]$ during time $T$. Namely, for all points $q \in [a,c]$, we consider external users who arrived during time \Cut{$[T(a),T(q)]$} \Col{$[0, T(q)]$}. Then ${\cal T}_{Q_0} ([a,c], T)$ is the sum of the service requirements of \RevCol{all} such users. We recall that $T(q)$ is a random quantity. \richardCol{Let ${\cal T}_{Q_j} ([a,c], T)$ denote the workload of the $j$-rerouted users in interval $[a,c]$.} \richardCol{${\cal T}_{Q_j} ([a,c], T)$ includes the fraction of the external arrivals ${\cal T}_{Q_0} ([0,|{\cal C}|], T)$ that have already received $j$ services and are now waiting in interval $[a,c]$ for their $(j+1)$-th service}.
Let
\begin{eqnarray}
 C_n^\sigma (q) \stackrel{\triangle}{=}  \phi_{n+1}^\sigma (q) - \phi_n^\sigma (q) = T_{n+1}^\sigma(q) + T_{n}^\sigma (|{\cal C}|) - T_{n}^\sigma (q) \hspace{-2mm} \label{Eqn_Cnq} \end{eqnarray}
represent the cycle  time w.r.t. $q$. 
 With this, (note at $\sigma = \infty$, $|{\cal C}|/\sigma = 0$):
 \begin{eqnarray}
 \label{Eqn_deltaTnq}
 T_n^\sigma(q) = T_n^\sigma(\delta^\sigma(q)) &=& \delta^\sigma (q) \alpha^{-1}  + {\cal T}_{Q_0} ([0, \delta^\sigma(q)), C_{n-1}^\sigma)  \nonumber \\
 && \hspace{14mm}
  + \sum_{j=1}^{N-1}  {\cal T}_{Q_j} \left ( \left [\frac{ |{\cal C}| }{\sigma}, \delta^\sigma(q)+ \frac{|{\cal C}|}{\sigma} \right ), C_{n-1-j}^\sigma \right ). \hspace{9mm}
 \end{eqnarray}
 This is the most important equation and is derived as follows.
 In (\ref{Eqn_deltaTnq}), the first term represents the time taken \RevCol{by the server to  travel the distance $\delta^\sigma(q)$}.
  The second term  represents the time taken to complete the service of the external arrivals at positions  before $\delta^\sigma(q)$. The third term represents
 the time taken to complete the service of the rerouted users, that arrived at positions between  $|{\cal C}|/\sigma$ and $\delta^\sigma(q)+|{\cal C}|/\sigma$.
 It is noted that $T_n^\sigma(q) = \phi_n^\sigma(0) - \phi_n^\sigma(q)$.
 Hence $T_n^\sigma(q)$ is the time period between the start of gated service at external queues at stop \kavfive{$\delta^\sigma(q)$ and the stop $0$.}  Therefore the time taken to \richardCol{serve} the rerouted users of the first stop, $\{ {\cal T}_{Q_j} ([0, |{\cal C}| /\sigma]) \}$, is not included in it. It instead includes
the time taken to serve the rerouted users of the  stop  $\delta^\sigma (q)$, $\{ {\cal T}_{Q_j} ([\delta^\sigma(q), \delta^\sigma(q) + |{\cal C}| /\sigma]) \}$.
{\it The $+$ \richardCol{symbol} in the third term is in the modulo $|{\cal C}| /\sigma$ sense. }

\subsection{First Moments}
\label{section_firstmoments}
We obtain an integral representation of the first moments of  ${\cal T}_{Q_0} ([0, q], T_n^\sigma)$, by extending 
\cite[Lemma 2]{JournPaper} to include mixed arrival position measures as in (\ref{Eqn_Arrivalprobs}).
\richardCol{
  Lemma \ref{Lemma1} is also true for open ($(a,c)$), semi-open ($(a, c]$, $[a,c)$) intervals and singletons ($\{a\}$). }
\begin{lemma}
\label{Lemma1}
Let $T:[a,c] \mapsto {\cal R}^+$ be either monotone (increasing or decreasing) or a constant nonnegative random function on the interval $[a,c]$. Assume that for any $q \in [a, c]$, the service times of the new arrivals and the arrival process at point $q$ are independent of the system evolution before time $T(q)$. Then (see equation (\ref{Eqn_Arrivalprobs})),
\begin{eqnarray*}
E[{\cal T}_{Q_0} ([a,c], T) ] \hspace{-2mm} &=& \hspace{-2mm} \lambda   \int_0^{|{\cal C}|}\hspace{-2mm} \indic_{\{[a,c]\} }(q) b_{0}(q)  \tau(q) P_{Q_0}(dq)  \mbox{ where } \tau(q) := E[T(q)] \forall q, \\
&& \hspace{-25mm} = \lambda \left ( \sum_{i=1}^{M_0} \indic_{\{ [a,c] \}}(q_{0,i})\tau(q_{0,i}) b_{0}(q_{0,i}) p_{0,i}  + \int_a^c b_{0}(q)  \tau(q) f_{Q_0}(q) dq \right ). \hspace{5mm} 
\end{eqnarray*}
\end{lemma}
 {\bf Proof :}
	Consider $A$ a Borel subset of $[0, |C|]$, and define ${\cal T}_{Q_0} (A, T)$ the number of arrivals in $A$ such that an arrival at $q \in A$ occurs during time $[0,T(q)]$.
	
	$\nu(A) := E[{\cal T}_{Q_0} (A, T)]$ defines a measure on the Borel subsets of $[0, |C|]$.
	We decompose the arrival process into two components, the arrivals due to the discrete component of $P_{Q_0}$, and the arrivals due to its absolutely continuous component and accordingly we will have $\nu = \nu_{disc}+ \nu_{cont}$.
Furthermore, using the Lebesgue decomposition theorem, we can write $\nu$ as a sum of a discrete measure, an absolutely continuous measure (with respect to the Lebesgue measure) and a singular measure $\nu =  \nu_{disc} + \nu_{leb} + \nu_{sing}$.

	\underline{Discrete component}
	
	For the discrete arrivals, we have that arrivals at point $a \in [0,|C|]$ during time $[0,T(a)]$ are independent of the system evolution before time $T(a)$. Furthermore the number of arrivals are independent of the service requirements. Hence by conditioning on $T(a)$ and applying Wald's lemma:	
	
	$$\nu_{disc}(A) = \sum_{i=1}^{M_{0}} \indic_{A}(q_{0,i}) b_{0}(q_{0,i}) \tau(q_{0,i}) p_{0,i} .$$
	
	\underline{Absolutely continuous component}

	If $A$ has Lebesgue measure $0$, the number of users arriving in $A$ due to the continuous  component of $P_{Q_0}$ during any arbitrary time is $0$ \ac{a.s}. Hence $\nu_{cont} = \nu - \nu_{disc}$ is absolutely continuous with respect to the Lebesgue measure, and $\nu_{sing} \equiv 0$.
		
	We first consider $T$ to be constant \ac{a.s}, $T(q) = T(a)$, $q \in [a,c]$ and we define $\tau = E[T(q)]$. Using the independence between the arrival process in $[a,c]$ during time $[0,T(a)]$ and the system evolution before time $T(a)$, the mean number of such arrivals is $\tau \lambda \int_a^c f_{Q_0}(q) dq$. The mean service times conditioned on the event that the arrival occurred in $[a,c]$ is equal to: $$\frac{\int_a^c b_{0}(q) f_{Q_0}(q) dq}{\int_a^c f_{Q_0}(q) dq} .$$
	
	 Using the independence between the arrival process and the service times we can apply Wald's Lemma to obtain:
	$$\nu_{leb}([a,c]) = \lambda \tau \int_a^c b_{0}(q) f_{Q_0}(q) dq.$$
	
	We now consider $T$ to be a non-decreasing random function. Let $\tau(q) = E[T(q)]$, and $q \mapsto T(q)$ is non-decreasing \ac{a.s}, hence $q \mapsto \tau(q)$ is non-decreasing. Define $Z_1$ the set of discontinuities of $\tau$, which is of Lebesgue measure $0$. Let 
\Col{$$Z_2 = \left \{ q_0 \in [0,|C|] \left | \limu{q_1}{q_0} \frac{1}{q_1 - q_0} \int_{q_0}^{q_1} b_{0}(q) f_{Q_0}(q)dq \neq b_{0}(q_0) f_{Q_0}(q_0) \right . \right  \},$$} and $Z_2$ is of Lebesgue measure $0$ by the Lebesgue differentiation theorem. 
	
	Consider \Cut{$Q_0 \notin Z = Z_1 \cup Z_2$, and $Q_1 > Q_0$} \Col{$q_0 \notin Z = Z_1 \cup Z_2$, and $q_1 > q_0$}. By monotonicity of $T$ we have that \Col{$T(q_0) \leq T(q) \leq T(q_1)$} and hence one can  bound  $\nu_{leb}$ appropriately as below.
We consider the arrivals in constant intervals \Col{$T(q_0)$, $T(q_1)$} in place of $T$ to obtain the lower and upper bounds respectively. 
 We can apply the previous result for $T$ constant \ac{a.s}, since the independence hypothesis is valid for the lower bound $T(q_0)$. The upper bound is obtained by considering independent arrivals that would have arrived in $[0, T(q_1)]$:
	$$ \tau(q_0) \lambda \int_{q_0}^{q_1} b_{0}(q) f_{Q_0}(q)dq \leq \nu_{leb}([q_0, q_1]) \leq \lambda \tau(q_1) \int_{q_0}^{q_1} b_{0}(q) f_{Q_0}(q)dq.$$
	
	Hence $\frac{1}{ q_1 - q_0} \nu_{leb}([q_0, q_1]) \tends{q_1}{q_0} \lambda \tau(q_0) b_{0}(q_0) f_{Q_0}(q_0) $ almost everywhere. \Cut{It is noted that we have not assumed $f_{Q_0}$ or $b_{0}$ continuous.}
	
	In particular for $A = [a,c]$, summing the discrete and absolutely continuous parts: 
	
	$$ E[{\cal T}_{Q_0} ([a,c], T)] = \lambda \left(\int_a^c b_{0}(q) \tau(q)f_{Q_0}(q) dq +  \sum_{i=1}^{M_{0}} b_{0}(q_{0,i}) \tau(q_{0,i}) p_{0,i}  \indic_{[a,c]}(q_{0,i}) \right)$$
	which concludes the demonstration. $\Box$

Similarly  for rerouted users   (proof in Appendix A):
\begin{lemma}
\label{LemmaRe}
With $T$ as in Lemma \ref{Lemma1},
the expected workload due to $j$-rerouted users equals (with $\tau(q) := E[T(q)]$) for $j > 1$:
\begin{eqnarray*}
E\left [{\cal T}_{Q_j}([a,c], T) \right ]
&=&   \hat{\epsilon}_j   \rho_{j} ([a,c])     \int_0^{|{\cal C}|} \tau(q) P_{Q_0}(dq) \mbox{ with } \hat{\epsilon}_j := \Pi_{i=1}^j \epsilon_i
. \ \ \Box
\end{eqnarray*}
\end{lemma}

We define $\tau^\sigma_n (q) := E[T_n^\sigma (q)]$ the first moment of $T_n^\sigma (q)$ for the discretized system and $\tau^\infty_n(q)$
\richardCol{the corresponding quantity} for the \cts system. Similarly we define $c^\sigma_n(q) := E [C_n^\sigma (q)]$.

\subsubsection{ Palm Stationarity and Stationarity:}  
 \label{footnote_palm}
 \richardCol{We recall the notion of Palm probability which is extensively used in this paper. More details are given in Appendix C.}
 For any stationary point process, such as $\{ \phi^\sigma_n(0) \}$, there are two associated probabilities: Stationary and Palm Stationary (\cite{Palm}). We use the convention that $\phi^\sigma_0 (0)  \le 0 <  \phi^\sigma_1 (0) $. 
Palm probability is the probability measure obtained by conditioning on the event $\{ \phi^\sigma_0 (0) = 0\}$ (see \cite{Palm} and Appendix C).
Throughout the paper, the Palm expectation is represented by $E^0$ and the corresponding moments are denoted with a ${.}_{*}$  \richardCol{subscript}.
In \cite{Palm},  stationary expectation of the residual cycle $C^\sigma_1 (0)$  (which we refer to as $C^\sigma_R(0)$) as well as the past cycle $C^\sigma_0 (0)$ (which we refer to as $C^\sigma_P(0)$) are obtained in terms of their two first Palm moments as
\begin{eqnarray}
\label{Eqn_ResCyc}
 E[C^\sigma_R (0)] = E[C^\sigma_P(0)] =  \frac{E^0\left [\left (C^\sigma_1 \right )^2(0) \right ]}{2 E^0[C^\sigma_1(0)]}.
 \end{eqnarray}
 \richardCol{A special case of this result is derived in \cite{Palm1}[Section 3.1]}. In Appendix C, we \richardCol{give} a brief summary of Palm probability and derive the above relation as equation (\ref{eq:palm_moment_3}). 

In \richardCol{the} Palm stationary regime, the distribution of $(\tau_n^\sigma, c_n^\sigma)$ does not depend on $n$, and they will be denoted $\tau^\sigma_*$, $c^\sigma_*$. Their distribution is the fixed point of an operator ${\cal F}$ given by equation (\ref{Eqn_taun}) at the end of this section. The fixed point gives their Palm moments. \richardCol{Using equation (\ref{Eqn_ResCyc}), we obtain the corresponding stationary moments in subsection~\ref{section_workload}.}

\mynCol{
\subsubsection{Fixed point equation}}

It is noted that that the cycle time $C_n^\sigma (q)$  is the  sum of monotone increasing ($q \mapsto T_n^\sigma(q)$)
and decreasing  ($q \mapsto T_{n-1}^\sigma(|{\cal C}|) - T_{n-1}^\sigma(q)$)  random functions of $q \in [0, |{\cal C}|].$ \richardCol{The workload is an additive function so by Lemmas \ref{Lemma1} and \ref{LemmaRe}}:
\begin{eqnarray}
\label{Eqn_taun}
\tau_n^\sigma(q)   &= &  \delta^\sigma (q) \alpha^{-1}  +
\lambda \int_0^{|{\cal C}|}  1_{\left\{[0,\delta^\sigma(q))\right \}} b_{0}(y)   c_{n-1}^\sigma(y) P_{Q_0}(dy)    \\
 && \hspace{2mm} + \sum_{j=1}^{N-1} \hat{\epsilon}_j  \rho_{j} \left ( \left [\frac{|{\cal C}|}{\sigma}, \delta^\sigma(q) +\frac{|{\cal C}|}{\sigma}\right ) \right )
  \int_0^{|{\cal C}|}     c_{n-1-j}^\sigma (y)  P_{Q_0}(dy)  \ \ \mbox{ and} \nonumber \\
c_n^\sigma (q) &=& \tau_n^\sigma(q) + \tau_{n-1}^\sigma\myCol{(|{\cal C}|)} -  \tau_{n-1}^\sigma (q) .
\nonumber
\end{eqnarray}

Let ${\cal D}$ \richardCol{denote} the Banach space of left continuous functions with right limits on  $[0, |{\cal C}|]$ equipped with the supremum norm,
$  ||\tau||_\infty :=  \sup_{q \in [0, |{\cal C}|]} |\tau(q)|. $
Let ${\mathbf N} := \{1, 2, \cdots, \infty\}$. 
 Define  function ${\cal F}$, parametrized by  $\sigma \in {\mathbf N}$, where ${\cal F}  (\tau; \sigma)$  is defined for any $\tau \in {\cal D}$ by:
\begin{eqnarray*}
{\cal F} (\tau ;  \sigma)  (q) &:=& \delta^\sigma (q) \alpha^{-1}  +  \tau (|{\cal C}|) \rho_{0} ([0, {\delta^\sigma(q)}  ))  
 \\ && 
  +   \tau (|{\cal C}|)    \sum_{j=1}^{N-1} \hat{\epsilon}_j  \rho_{j} \left ( \left [\frac{|{\cal C}|}{\sigma}, \delta^\sigma(q) +\frac{|{\cal C}|}{\sigma}\right ) \right )
 \mbox{for all $q \in [0, |{\cal C}|]$  .}
\end{eqnarray*} 
Let  $\tau^\sigma_* (q), c^\sigma_*(q) $, for $\sigma \in {\mathbf N}$, \richardCol{denote the stationary quantities corresponding to the Palm moments} $\tau_n^\sigma(q)$, $c_n^\sigma(q)$ respectively. By {\it Palm stationarity} and from equation (\ref{Eqn_taun}), the stationary first moments of the discrete system are the fixed points of the parametrized function ${\cal F}$, at $\sigma < \infty$ while \richardCol{the stationary first moments of} the \cts system are the fixed points of the same function at $\sigma = \infty.$  

The only limit point of  the set ${\mathbf N}$  (in Euclidean metric) is  $\infty$.
Thus the function ${\cal F}$ is continuous in $(\tau, \sigma)$  (with  $\tau \in {\cal D}$, $\sigma \in {\mathbf N}$) because:
1) ${\cal F}$ is bounded linear in $\tau$;  2) as $\sigma \to \infty$,
 $\delta^\sigma \to \delta^{\infty}$  in sup norm    and ${\cal F}$ is continuous
in $\delta \in {\cal D} $  (when ${\cal F}$ is viewed as a function of $\tau, \delta, \sigma$ after replacing $\delta^\sigma$ with $\delta$).
Hence  (see   proof of  \cite[Theorem 2]{JournPaper} for \mynCut{some more} \mynCol{similar} details) using the contraction mapping theorem:
\begin{thm}
\label{Thrm_FirstMoments}
For any $\sigma$,
 the  map ${\cal F}$ has an unique fixed point, $\tau^\sigma_*$, if and only if
$\rho = \rho_{0} +  \sum_{j=1}^{N-1} {\hat \epsilon}_j  \rho_{j} < 1. $
  Furthermore, the stationary moments of the discrete system $\tau^\sigma_*$ converge to \richardCol{the stationary moments} of the \cts system:
\[
 \sup_{ q \in {\cal C}} | \tau^\sigma_* (q) - \tau^\infty_* (q) |  \to 0   \mbox{ as }  \sigma \to \infty.
\]
Indeed \richardCol{using} modulo  $|{\cal C}| /\sigma$  addition for all $q$
\begin{eqnarray*}
\tau^\sigma_* (|{\cal C}|)  =
 \frac{\delta^\sigma(|{\cal C}| )}{\alpha(1 - \rho)}  \mbox{ and } \hspace{85mm} \\
\tau^\sigma_*(q) = \frac{\delta^\sigma(q)}{ \alpha}  + \tau^\sigma_*(|{\cal C}|) \left ( {\rho_{0}} \left (\left[ 0, \delta^\sigma(q)\right ) \right) 
+   \sum_{j=1}^{N-1} {\hat \epsilon}_j  \rho_{j}   \left ( \left [\frac{|{\cal C}|}{\sigma}, \delta^\sigma(q) +\frac{|{\cal C}|}{\sigma}\right ) \right ) \right ).
\mbox{  } \Box
\end{eqnarray*}
\end{thm}

\subsection{Second Moments}

\richardCol{We define $c_{*}^{(\sigma 2)}(q)$, the (Palm) stationary second moments of $C_{n}^{\sigma}(q)$. The second moments for the \cts polling system are $c_{*}^{(\infty 2)}(q)$.}
We have the following:
\begin{thm}
\label{Thrm_SecondMoments}
There exists a threshold $\bar{\pi}_b^{(2)} > 0$ on the second moments \richardCol{of} the service times such that if ${\bar b}^{(2)}_{j} <
\bar{\pi}_b^{(2)}$ for all $j$, then the stationary second moments of the discretized system converge to \richardCol{the stationary second moments} of the \cts system. \richardCol{Convergence occurs along a subsequence $\sigma_k \to \infty$ almost surely with respect to $Q_0$.} Namely:
\begin{eqnarray*}
   P_{Q_0}\left ( \left | c_{*}^{(\sigma_k 2)}(Q_0) -  c_{*}^{(\infty 2)}(Q_0) \right |  
 \tends{\sigma_k}{\infty} \ \ 0  \right ) = 1 . 
\end{eqnarray*}
Furthermore there exists a ${\bar \sigma} < +\infty$ and a constant ${\bar \omega} < +\infty$ such that 
$$
c_{*}^{(\sigma_k 2)}(q) \le {\bar \omega} \mbox{ for all  }q, \mbox{ and } \sigma_k > {\bar \sigma}.  \hspace{3mm} \Box
$$
\end{thm}

\subsection{Final Step}
\label{section_workload}

By Theorems \ref{Thrm_FirstMoments}  and \ref{Thrm_SecondMoments}, the first stationary moments of $T_n^\sigma$ and the  second  stationary moments of $C_n^\sigma$
of the discretized polling system
converge towards the  corresponding \richardCol{moments} of the \cts polling system. \richardCol{For the second moments we obtained this convergence along a subsequence and it suffices to work with that subsequence.} {\it To simplify the notations, we represent the subsequence again by $\sigma \to \infty$.} 
We obtain a common expression for the expected virtual workload using the two \Col{sets} of moments and complete the proof.


From  (\ref{Eqn_Cnq}), the first two (Palm) stationary moments of the cycle time w.r.t. the point $q$, $E^0[C^\sigma_n(q)]$ and $E^0\left[ (C^\sigma_n(q))^2 \right] $  are:
{
\begin{eqnarray}
E^0[C^\sigma_n(q)] =: c_*^\sigma (q) = \tau^\sigma_* (|{\cal C}|)  \nonumber  \mbox{ and }
E^0\left[ (C^\sigma_n(q))^2 \right]  = c_*^{(\sigma 2)} (q).
\end{eqnarray} }
\richardCol{Hence} the first stationary moment of the residual of the cycle  $C^\sigma_n(q)$ as seen by a random user is given by (see  equation (\ref{Eqn_ResCyc}) of \myCol{subsection} \ref{footnote_palm}),
\begin{eqnarray}
E [C_R^\sigma (q) ]  &=& \frac{ c_*^{(\sigma 2)} (q)  }{2 c_*^\sigma (q)}. \label{Eqn_ResCycles}
\end{eqnarray}

In the following we calculate the expected \richardCol{stationary} workload due to external users and rerouted users separately, in terms of the stationary moments of the cycle times studied in previous subsection,
using Little's law and then the Wald's Lemma.

\noindent \underline{\bf Workload due to external users:} \hspace{3mm}
An external user arriving at $q$ has to wait for:
\begin{enumerate}
\item the residual time of his own cycle $C_*^\sigma(q)$;
\item the time taken to \richardCol{serve} the external users waiting at $q$ that arrived before him (FIFO);
\item  in a discrete system (arrival position order service), \richardCol{the time} until the external users waiting in $[\delta^\sigma(q), q)$)  are served.
\end{enumerate}
\Col{
We are interested in the average of this waiting time. The average of the 
 total time due to points 2 and 3 is given by}
  $E \left [ {\cal T}_{Q_0} \left( \left [\delta^\sigma(q), q \right ],{\ddot C}  \right ) \right ]$, where\footnote{The users of point 2 arrive at the same point $q$ before the tagged user and within the current cycle, i.e. during the past cycle $C_P^\sigma(q)$. 
}
\begin{eqnarray*}
{\ddot C}(q') :=  \left \{ \begin{array}{llllll}
C_*^\sigma(q')   & \mbox{if } q'\ne q  &\mbox{- stationary cycle corresponding to $C_n(q')$ and } \\
 C_P^\sigma (q) & \mbox{if } q' = q  &\mbox{- past cycle as in   subsection \ref{footnote_palm}. }  \end{array} \right . \end{eqnarray*}
 By Lemma \ref{Lemma1} (see (\ref{Eqn_Arrivalprobs}), note $c_*^\sigma (q') = E[C_*^\sigma(q')] =\tau_*^\sigma (|{\cal C}|)$ for all $q'$ 
 and  $E[C_R^\sigma(q)] = E[C_P^\sigma(q)]$),
{\small \begin{eqnarray}
E \left [ {\cal T}_{Q_0} \left( \left [\delta^\sigma(q), q \right ], {\ddot C}  \right ) \right ] =
  \tau^\sigma_*(|{\cal C}|) \rho_{0} \left( \left [\delta^\sigma(q), q \right )  \right )
  + \sum_{i=1}^{M_0} p_{0,i} 1_{\{q_{0,i}\}} (q) E[C_R^\sigma(q)] .
\label{Eqn_UserdueRes}
\end{eqnarray} }
Thus the expected waiting time of an external user is:
\begin{eqnarray}
\label{Eqn_WaitTim}
E [W_{Q_0}^\sigma] (q) = E [C_R^\sigma (q) ]
+
E \left [ {\cal T}_{Q_0} \left( \left [\delta^\sigma(q), q \right ], {\ddot C}  \right ) \right ].
\end{eqnarray}
\richardCol{The expected waiting time of an external user in} a \cts system  equals:
\begin{eqnarray}
\label{Eqn_WaitTimInf}
E [W_{Q_0}^\infty] (q) = E [C_R^\infty (q) ]
  + \sum_{i=1}^{M_0} p_{0,i} 1_{\{q_{0,i}\}} (q) E[C_R^\infty(q)] .
\end{eqnarray} 
By  Little's law  (see Appendix C), the stationary expected number of users waiting \richardCol{for} their first service in an infinitesimal interval $[q-dq, q+dq]$ equals $\lambda    E [W_{Q_0}^\sigma] (q) P_{Q_0}(dq) $. By Wald's Lemma, the stationary expected virtual workload due to these external users equals  $\lambda   E [W_{Q_0}^\sigma] (q) b_{0}(q) P_{Q_0}(dq)$ by independence.
Thus the expected stationary workload due to all the external users  is:
\[
{V}_{Q_0}^\sigma = \int_0^{|\cal C|} \lambda    E[W_{Q_0}^\sigma] (q)    b_{0}(q) P_{Q_0}(dq) .
\]
By Lemma~\ref{Lemma_fI} (Appendix B)  $\rho_{0} \left( \left [\delta^\sigma(q), q \right )  \right ) \to 0$ as $\sigma \to \infty$. 
\richardCol{Using the Bounded Convergence Theorem} (BCT) and Theorems \ref{Thrm_FirstMoments}, \ref{Thrm_SecondMoments} we have that 
$E [C_R^\sigma (q) ] \to E [C_R^\infty (q) ] $  
and hence $E[W_{Q_0}^\sigma] (q)   \to  E[W_{Q_0}^\infty] (q) $  as $\sigma \to \infty$ for every $q$ (see equations (\ref{Eqn_ResCycles})-(\ref{Eqn_WaitTimInf})). 
\richardCol{Applying the BCT} again, ${V}_{Q_0}^\sigma \to {V}_{Q_0}^\infty.$

\noindent \underline{\bf  Workload due to rerouted users ($j>0$):} 
\richardCol{Because of immediate rerouting, a $j$-rerouted user arrives just after completion of his $j$-th service to another point on the circle.
Consider such a user arriving at point $q$.}  
His waiting time depends on where his $j$-th service was \richardCol{received}. If he received the $j$-th service at $q'$ (whose
distribution is  given by $P_{Q_{(j-1)}} (dq')$ because of independence), he will  have to wait on average for: 
\begin{enumerate}
\item if $q ' < q$ : a period of time $\tau^\sigma_*(q) - \tau^\sigma_*(q')$ ;
\item  if $q' > q$ : a period of time  $\tau^\sigma_*(|{\cal C}|) - \tau^\sigma_*(q') + \tau^\sigma_*(q)$.  
\end{enumerate}
The above \richardCol{quantity is} the waiting time until the external arrivals of interval $[\delta^\sigma(q), \delta^\sigma(q)+|{\cal C}|/\sigma)$ are served.
The rerouted user will be served \underline{before} this time. Let $\nu_j^\sigma(q)$ \richardCol{denote} the average of this time difference.
This difference is calculated and its convergence is studied for all the polling systems in Lemma \ref{Lemma_TimDiff} of Appendix B.
Then the stationary average waiting time of a $j$-rerouted user arriving at point $q$ equals:
{\small
\begin{eqnarray*}
E[W_{Q_{j}}^\sigma] (q) &=&
   \int_q^{|{\cal C}|}   \tau^\sigma_*(|{\cal C}|)  P_{Q_{(j-1)}} (dq')
   +  \int_0^{|{\cal C}|} \hspace{-2mm} \left (\tau^\sigma_*(q) - \tau^\sigma_*(q') \right )  P_{Q_{(j-1)}} (dq')
   -\nu_j^\sigma(q) \\
&=& 
   \tau^\sigma_*(q) + \tau^\sigma_* (|{\cal C}|) P_{Q_{(j-1)}} ([q, |{\cal C}|]) -  \int_0^{|{\cal C}|}  \tau^\sigma_*(q')  P_{Q_{(j-1)}} (dq')
   -\nu_j^\sigma(q)
\end{eqnarray*} }
By  Little's law  (see Appendix C) and Wald's lemma, as before,   the stationary expected  workload due to $j$-rerouted users
is:
$$
V_{Q_{j}}^\sigma = \lambda {\hat \epsilon}_j  \int_0^{|{\cal C}|}  E[W_{Q_{j}}^\sigma] (q)    b_{{j}} (q) P_{Q_j} (dq) .
$$
It is noted that the effective arrival rate equals ${\hat \epsilon}_j \lambda P_{Q_j} (dq)$. By theorems \ref{Thrm_FirstMoments}, \ref{Thrm_SecondMoments} and \richardCol{the} BCT  (as for the external users) and Lemma \ref{Lemma_TimDiff} of Appendix B, we obtain: 
 ${V}_{Q_{j}}^\sigma \to {V}_{Q_{j}}^\infty.$

\noindent\underline{\bf Total workload:}
The total expected stationary virtual workload  is the workload due to all types of users and hence
$
V^\sigma =   V_{Q_0}^\sigma + \sum_{j=1}^{N-1} V_{Q_{j}}^\sigma.
$ From the above arguments, $V^\sigma \to V^\infty$. Namely the
total stationary expected virtual workload of the discrete polling system converges \richardCol{to the total stationary expected virtual workload} of the \cts polling system as $\sigma \to \infty.$
The workload $V^\sigma$ is obtained as an easily computable expression for each $\sigma$ in subsection \ref{section_discrete} using results of \cite{SLF} as  equation (\ref{Eqn_MixedService}). \richardCol{Its} limit  (\ref{Eqn_Vmix}) \richardCol{is} the stationary expected  workload of the \cts system.  This completes the proof of Theorem 1. $\Box$




\section{Conclusion}
\label{section_conclusions}
We study a \cts polling system in which 
\Col{the users can either arrive in a continuum or queue up at \myCol{a finite number of locations}.
These systems generalize both discrete and continuous polling systems.
\Col{The defining feature of \cts polling systems is that their arrival position measure is a mixture of discrete and continuous probability measures.}
\richardColNew{Furthermore, the model allows rerouting:} users can be rerouted to a series of independent positions to receive multiple services. The number of services received is a random number, upper bounded by a constant integer almost surely. We obtain \Col{the} stationary performance of a \cts system with rerouting under more general conditions than the usual symmetric conditions.}

\Col
{
We adapted the previously used discretization approach (\cite{JournPaper}) to our \cts polling system. Because of the existence of atoms in the arrival position measure (discrete parts in the probability measure) 
we had to use "weaker" notions of convergence. In \cite{JournPaper} uniform convergence of certain stationary moments is obtained at almost all the points on the cyclic path while we managed to obtain convergence only in $L^1$ norm w.r.t. the arrival position measure on the cyclic path. However we still \richardColNew{manage} to obtain the \Col{main} result: 
  the expected workload of a \cts polling system is the limit of the expected workloads of a sequence of discrete systems with rerouting for which computable expressions are available.
}

\Col{
We apply our results to a wireless LAN in which a ferry \richardCol{enables} data transfer between users of a network who are not directly connected, while moving in a rectangular path. The rectangular path \Col{results in a mixed arrival position measure} (queues are formed \Col{at} the corner points) in the equivalent polling system. Thus this example could not be modeled with any of the \myCol{conventional} polling systems. Data transfer also occurs between users and the outside world using a gateway. We further consider a hybrid architecture in which the ferry is \richardCol{assisted} by a node called base station which manages the reroutings. We study the optimal system design using numerical examples. We propose some design guidelines based on the results. For instance at large ferry speeds, the ferry path optimizing the first moment of the service time is close to the one optimizing the workload. With large ferry speeds or with long ferry routes, the optimal rectangular ferry path divides the breadth of the network area into equal partitions. 
The optimal ferry path does not \Col{change} much whether or not a base station is involved in \Col{the} local data transfer. }

We also apply our results to a waste collector where waste is generated randomly over a large area. A dedicated vehicle must pick it up and discard it to a collection point. We study the optimal location of the collection point given the cumulative distribution $F_Q (.)$ of the locations of waste generation. The {workload} performance does not depend on the \richardCol{collection point} if the waste is generated uniformly on the area while in a general case, the optimal \richardCol{collection point}  minimizes  the function, $x - F_Q(x)$. 

\section*{Appendix A}
{\noindent}{\bf Proof of  Lemma \ref{Lemma_LimVmixsigma}:}

The first three terms of the r.h.s of (\ref{Eqn_DiscreteCmbinedMain}) are independent of $\sigma$ and simplify as the first
\richardCol{three} terms of (\ref{Eqn_Vmix}). 

We recall that we use the circular sum convention (see \cite{SLF}) so that :
\begin{align*}
	\sum_{k = Nl}^{Ni} {.} &=  \sum_{k=Nl}^{N\sigma} {.}   +  \sum_{k=1}^{Ni} {.}    \text{ if } l > i.
\end{align*}
Hence:
\begin{align*}
\sum_{k = Nl}^{Ni} \rho_k &= 1_{\{ i \ge l \} } \rho \left( \left[ l^\sigma, i^\sigma \right ) \right)
+  1_{\{ i < l \} }  \left [ \rho -  \rho \left  ( \left[ i^\sigma, l^\sigma \right )  \right ) \right ] \\
  &= \rho ([0, i^\sigma)) -  \rho([0, l^\sigma)) + 1_{\{ i < l \} }  \rho  \\
  &= {\hat \rho} ( i^\sigma) -  {\hat \rho}( l^\sigma) + 1_{\{ i < l \} }  \rho.
\end{align*}
We define the function $g_0^\sigma$ on ${ [0,|{ \cal C}|] }^2$:
\begin{equation}\label{eq:f1}
	g_0^\sigma(q,q^\prime) = \frac{\alpha^{-1}}{1 - \rho } \left( b_{0}(q) +  \sum_{j=1}^{N-1} {\hat \epsilon}_j {\bar b}_{j} \right ) (  {\hat \rho} (i^\sigma) -  {\hat \rho}( l^\sigma) + 1_{\{ i^\sigma < l^\sigma \} } \rho )  , 
\end{equation}
if $(q,q^\prime) \in I_l \times I_i$, and the measure $m_0(dq \times dq^\prime) = P_{Q_0}(dq) dq^\prime$. \richardCol{Then} $g_0^\sigma$ is bounded on ${ [0,|{ \cal C}|] }^2$, and converges $m_0$-almost everywhere:
\begin{equation}\label{eq:f1c}
g_0^\sigma(q,q^\prime) \tends{\sigma}{+\infty} g_0(q,q^\prime) =  \frac{\alpha^{-1}}{1 - \rho } \left( b_{0}(q) +  \sum_{j=1}^{N-1} {\hat \epsilon}_j {\bar b}_{j} \right ) (  {\hat \rho} ( q^\prime) -  {\hat \rho}( q) + 1_{\{ q^\prime < q \} }  \rho).
\end{equation}
The fourth term of the r.h.s of (\ref{Eqn_MixedService}) is equal to:
\begin{align}
\int_{0}^{|{\cal C}|} \int_{0}^{|{\cal C}|} g_0^\sigma(q,q^\prime) dm_0(dq \times dq^\prime) \tends{\sigma}{+\infty} \int_{0}^{|{\cal C}|} \int_{0}^{|{\cal C}|} g_0(q,q^\prime) dm_0(dq \times dq^\prime)  \nonumber\\
= \frac{\alpha^{-1}}{1 - \rho } \int_{0}^{|{\cal C}|} \int_{0}^{|{\cal C}|}   \left( b_{0}(q) +  \sum_{j=1}^{N-1} {\hat \epsilon}_j {\bar b}_{j} \right ) ( {\hat \rho} ( q^\prime) -  {\hat \rho}( q) + 1_{\{ q^\prime < q \} }  \rho) P_{Q_0}(dq) dq^\prime.
\label{Eqn_approx}
\end{align}
\richardCol{by the dominated convergence theorem}. We \richardCol{consider} the $j$-th term of the fifth term of (\ref{Eqn_MixedService}). The circular sum equals:
\begin{eqnarray*}
\sum_{k = Nl}^{Ni + j} r_k =  \alpha^{-1} ( i^\sigma - l^\sigma  + |{\cal C}|1_{\{ i^\sigma < l^\sigma \} } ).
\end{eqnarray*}
because $r_{k} = \alpha^{-1} |{\cal C}|/\sigma$ if $k$ is a multiple of $N$, and $0$ otherwise. Define function $g_j^\sigma$:
\begin{equation}\label{eq:fj}
	g_j^\sigma(q,q^\prime) = \frac{ \hat{\epsilon}_{j+1} \lambda \alpha^{-1}}{1 - \rho }  \left ( b_{{j+1}} (q^\prime) +  \sum_{k=j+2} \check{\epsilon}_{j+2}^k {\bar b}_{k} \right ) (  i^\sigma - l^\sigma  + |{\cal C}| 1_{\{ i^\sigma < l^\sigma \} } ),
\end{equation}
if $(q,q^\prime) \in I_i \times I_l$, and measure $m_j(dq \times dq^\prime) = P_{Q_j}(dq) P_{Q_{j+1}}(dq^\prime)$. Then $g_j^\sigma$ is bounded on $[0,|{\cal C}|]^2$, and it converges $m_j$ - almost everywhere:
\begin{equation}\label{eq:fjc}	
 g_j^\sigma(q,q^\prime) \tends{\sigma}{+\infty}  g_j(q,q^\prime) =  \frac{ \hat{\epsilon}_{j+1} \lambda \alpha^{-1}}{1 - \rho } \left ( b_{{j+1}} (q^\prime) +  \sum_{k=j+2} \check{\epsilon}_{j+2}^k {\bar b}_{k} \right ) (  q^\prime - q  + |{\cal C}| 1_{\{ q > q^\prime \} } )
	\end{equation}
	It is noted that $g_j^\sigma(q,q) = g_j(q,q) = 0$ $\forall \sigma$ so that the convergence \richardCol{holds} despite the fact that $(q,q^\prime) \to  1_{\{ q < q^\prime \} }$ is not continuous at $(q,q)$.
Then the $j$-th term of the fifth term of (\ref{Eqn_MixedService}) is:
\begin{align*}
 \int_{0}^{|{\cal C}|} & \int_{0}^{|{\cal C}|} g_j^\sigma(q,q^\prime) dm_j(dq \times dq^\prime) \tends{\sigma}{+\infty} \int_{0}^{|{\cal C}|} \int_{0}^{|{\cal C}|} g_j(q,q^\prime) dm_j(dq \times dq^\prime) \\
	=& \frac{ \hat{\epsilon}_{j+1} \lambda \alpha^{-1}}{1 - \rho } \int_{0}^{|{\cal C}|} \int_{0}^{|{\cal C}|}  \left ( b_{{j+1}} (q^\prime) +  \sum_{k=j+2} \check{\epsilon}_{j+2}^k {\bar b}_{k} \right ) ( q^\prime - q  + |{\cal C}| 1_{\{ q > q^\prime \} } ) P_{Q_j}(dq) P_{Q_{j+1}}(dq^\prime),
\end{align*}
\richardCol{by the dominated convergence theorem}, which concludes the proof. \hfill{$\Box$}

\medskip
\medskip

\noindent{\bf Proof  of  Lemma \ref{LemmaRe} :}
Let ${\cal N}$ \richardCol{denote} the number of users that caused the workload ${\cal T}_{Q_0} ([0, |{\cal C}|], T)$. 
We have that ${\cal N}  = {\hat {\cal T}}_{Q_0} ([0, |{\cal C}|], T)$,  where $ {\hat {\cal T}}_{Q_0} ([0, |{\cal C}|], T)$ \richardCol{denotes} the workload due to the arrivals which request unit service time (i.e. assuming $B_{Q_0} \equiv 1$, so $b_{0}(q) =1$ for all $q$). Hence by Lemma \ref{Lemma1}
\[
E[{\cal N}] = \lambda   \int_0^{|{\cal C}|} f_{Q_0}(q) \tau(q) dq \mbox{ where } \tau(q) = E[T(q)] .
\]
Let ${\cal N}_{Q_j}$ \richardCol{denote} the number of  $j$-rerouted users among  these ${\cal N}$. Then (by independence of rerouting locations)
\begin{eqnarray}
\label{Eqn_Nr}
{\cal N}_{Q_j} = \sum_{i=1}^{\cal N} 1_{\left \{\mbox{user $i$ was $j$-rerouted to $[a,c]$} \right \}}.  \hspace{2mm}
\end{eqnarray} The user is rerouted after every service, and the rerouting location is independent of the system evolution up to that time. Hence by applying Wald's lemma to (\ref{Eqn_Nr}):
$$
E[{\cal N}_{Q_j}]  = E[{\cal N}] E\left [1_{\left \{\mbox{$j$-reroute to } [a,c] \right  \}}\right ] = E[{\cal N}] \Pi_{i=1}^j \epsilon_i P_{Q_j} ([a,c]).
$$
The service time requirement in the  $(j+1)$-service $B_{Q_j}$ is independent of the system evolution up to that time. Its expected value conditioned on the fact that the user arrival is in interval
$[a,c]$ is $E[B_{Q_j} |  Q_j \in [a,c]]$. Hence
by applying Wald's lemma again:
$$ 
\hspace{20mm}E[{\cal T}_{Q_j}([a,c], T)] ={\hat \epsilon}_j P_{Q_j} ([a,c]) E[B_{Q_j} |  Q_j \in [a,c]]  E[{\cal N}].   \mbox{\hfill{$\hspace{21mm}\Box$} } $$

\noindent{\bf Proof of Theorem \ref{Thrm_SecondMoments}:}
	We rewrite the cycle times $C_n^\sigma$ from equations (\ref{Eqn_deltaTnq}) and (\ref{Eqn_Cnq}) as below (\richardCol{using} modulo $|{\cal C}|$ addition):
	\begin{eqnarray}
	\label{Eqn_Cnsigma}
 C_n^\sigma(q) &= & |{\cal C}| \alpha^{-1} + \sum_{j=0}^{N-1} U_{n,j}^\sigma (q) + {\cal T}_{Q_0} ( [0, \delta^\sigma(q)), C_{n}^\sigma )   \mbox{ with} \\
 U_{n,o}^\sigma(q) &:=&  {\cal T}_{Q_0} \left ( \left [ \delta^\sigma(q),  |{\cal C}| \right ), C_{n-1}^\sigma \right ) 
 +  {\cal T}_{Q_1}   \left ( \left [ \frac{ |{\cal C}| }{\sigma}, \delta^\sigma(q)+ \frac{|{\cal C}|}{\sigma} \right ), C_{n-1}^\sigma \right )  \nonumber \\
 U_{n,j}^\sigma(q) &:=&  {\cal T}_{Q_j} \left ( \left [ \delta^\sigma(q) + \frac{ |{\cal C}| }{\sigma},  
|{\cal C}| + \frac{ |{\cal C}| }{\sigma} \right ), C_{n-1-j}^\sigma \right )  \nonumber \\
&& \hspace{5mm}
 +  {\cal T}_{Q_{j+1}}   \left ( \left [ \frac{ |{\cal C}| }{\sigma}, \delta^\sigma(q)+ \frac{|{\cal C}|}{\sigma} \right ), C_{n-1-j}^\sigma \right ) \nonumber  \mbox{ for } 1 \le j \le N-2 \\
 U_{n,N-1}^\sigma(q) &:=&  {\cal T}_{Q_{N-1}} \left ( \left [ \delta^\sigma(q) + \frac{ |{\cal C}| }{\sigma},  
|{\cal C}| + \frac{ |{\cal C}| }{\sigma} \right ), C_{n-N}^\sigma \right ).  \label{Eqn_Unsigma}
 \end{eqnarray}
		
\underline{Function space}

\richardCol{Let $(\Omega,{\cal F},P)$ denote the probability space over which the entire system evolution (\richardCol{arrivals, reroutings and service times}) is defined.} Define $\phi:[0,|{\cal C}|] \times \Omega \to \RR$, and 
\Col{
$$\norm{\phi}_{1,2} =  E_{Q_0}[ ||\phi(Q_0)||_2], \mbox{ with } ||\phi(q)||_2 := \sqrt{E[\phi(q)^2]} = \sqrt{E_P [\phi (q, .)^2]}$$ the $L^2$ norm} (with respect to probability measure $P$) of the 
$q$-section of the function $\phi$. We recall that the expectation $E_{Q_0}$ is with respect to $Q_0$. Define 
\begin{equation}\Psi := \left \{ \phi:[0,|{\cal C}|] \times \Omega \to \RR \,|\, q \mapsto \sqrt{E[\phi(q)^2]} \text{ is c\`adl\`ag and }\norm{\phi}_{1,2} < +\infty \right \}.\end{equation} 
Then $\Psi$ equipped with $\norm{}_{1,2}$ is a normed linear  space. 
For the $n$-product space $\Psi^n := \Psi \times \Psi \cdots \Psi$, we define the norm:  
\vspace{-4mm}
 \begin{eqnarray}
 \label{Eqn_ProdNorm}
||( \phi_1, \cdots, \phi_n )||_{1,2}  := E \left [\sqrt {   \sum_{i=1}^{n} || \phi_{i} ||_2^2 } \right ]. 
\end{eqnarray}

\underline{Fixed point equations and contractivity}

Define $\pi_b^{(2)} := \max_j \sqrt{ {\bar b}_{j}^{(2)}}.$
	For any fixed $\sigma, \phi_c, \phi_u := (\phi_{u_0}, \cdots, \phi_{u_{N-1}})$, we define the \richardCol{mapping}, $\Theta :=(\Theta_c,\Theta_{u_0}, \cdots, \Theta_{u_{N-1}};\sigma): \Psi^{N+1} \to \Psi^{N+1}$ \Col{point-wise} by:	
\begin{eqnarray}
\label{Eqn_FixedC}
\Theta_c(\phi_c,\phi_u; \sigma)(q) &:= &|{\cal C}|\alpha^{-1} + \sqrt{\pi_b^{(2)}}\sum_{j=0}^{N-1} \phi_{u_j}(q) + {\cal T}_{Q_0} ([0,\delta^{\sigma}(q)), \phi_c)\\
	\Theta_{u_0}(\phi_c,\phi_u; \sigma)(q) &:=  & \frac{1}{\sqrt{\pi_b^{(2)}}}
{\cal T}_{Q_0} \left ( \left [ \delta^\sigma(q),  |{\cal C}| \right ), \phi_c \right )  \\ && \hspace{5mm}
 +  \frac{1}{\sqrt{\pi_b^{(2)}}} {\cal T}_{Q_1}   \left ( \left [ \frac{ |{\cal C}| }{\sigma}, \delta^\sigma(q)+ \frac{|{\cal C}|}{\sigma} \right ), \phi_c \right )  \nonumber \\
 \Theta_{u_j}(\phi_c,\phi_u;  \sigma)(q) &:=& \frac{1}{\sqrt{\pi_b^{(2)}}} {\cal T}_{Q_j} \left ( \left [ \delta^\sigma(q) + \frac{ |{\cal C}| }{\sigma},  
|{\cal C}| + \frac{ |{\cal C}| }{\sigma} \right ), \phi_c \right )  \nonumber \\
&& \hspace{5mm}
 + \frac{1}{\sqrt{\pi_b^{(2)}}} {\cal T}_{Q_{j+1}}   \left ( \left [ \frac{ |{\cal C}| }{\sigma}, \delta^\sigma(q)+ \frac{|{\cal C}|}{\sigma} \right ), \phi_c \right ) \nonumber \\ && \hspace{30mm} \mbox{ for all } 1 \le j \le N-2. \hspace{10mm} \\
 \Theta_{u_{N-1}}(\phi_c,\phi_u; \sigma)(q) &:=&  \frac{1}{\sqrt{\pi_b^{(2)}}} {\cal T}_{Q_{N-1}} \left ( \left [ \delta^\sigma(q) + \frac{ |{\cal C}| }{\sigma},  
|{\cal C}| + \frac{ |{\cal C}| }{\sigma} \right ),  \phi_c \right ).  \label{Eqn_FixedU}
\end{eqnarray}
The definition of the term ${\cal T}_{Q_j}$ is  extended naturally \richardCol{to} negative times $\{\phi_c(q)\}_q$. For any $j$, the term ${\cal T}_{Q_j} ([a,c], \phi_c)$ 
 \richardCol{denotes} the sum of the weighted service times of users that have arrived in (or have been rerouted to)  $[a,c]$ between times $0$ and $|\phi_c(q)|$, $q \in [a,c]$. The weight equals  $1$ if $\phi_c(q) \geq 0$ and  $-1$ otherwise. These service times are independent of the other random quantities.

 We are interested in the fixed points $(\phi^*_c, \phi^*_u)$  of the mapping $\Theta$ defined by the equations 
(\ref{Eqn_FixedC})-(\ref{Eqn_FixedU}). Indeed, if
\begin{itemize}
\item  $C_{n-j}^\sigma  \stackrel{distribution}{=}  \phi_c^*$ ,  $0\le j \le N$ and,
\item $U_{n,j}^\sigma  \stackrel{distribution}{=} \phi_{uj}^*$ for all $j$,
\end{itemize}
then $C^\sigma_n  \stackrel{distribution}{=} \phi_c^*$.
This is true for any $n$ (see  equations (\ref{Eqn_Cnsigma})-(\ref{Eqn_Unsigma})). Therefore the fixed points of $\Theta$ are the stationary quantities. Note here that $C_n^\sigma (q)$, for any $\sigma$ \Col{ and q}, depends upon $C_{n-j}^\sigma$ for all $ 1 \le j \le N$ and $C^\sigma_n (q') $ for all $q' < q$.
		
	$\Theta$ is an affine operator and we write $\Theta_{lin}$ its linear part.
	This linear operator \richardCol{involves terms like} ${\cal T}_{Q_j} ([a,b], \phi)$.  Each of these 
	terms can be written as \richardCol{a} sum of service time requirements, with ${\cal N}_{Q_j}$ the corresponding number of arrivals (note that ${\bar b}_{j}^{(2)} = {E[b_{j}^{(2)} (Q_j)]} = {E [B_{Q_j}^2] }$ ):
	\begin{equation} {\cal T}_{Q_j} (I,\phi)  = \sum_{i=1}^{{\cal N}_{Q_j}(I,\phi)} B_{Q_j} (i)  =  \sqrt{{\bar b}_{j}^{(2)}} \sum_{i=1}^{{\cal N}_{Q_j}(I,\phi)} {\tilde B}_{Q_j}(i) , \end{equation}
	where ${\tilde B}_{Q_j}(i)$ \richardCol{denotes} the service times with unit second moment. It is noted that the sign of $B_i$ or the number of reroutings do not matter for this discussion. We can see that for any interval $I$  and any $\phi$ and $j$:
	$$
	|| {\cal T}_{Q_j} (I, \phi) ||_2 =   \sqrt{{\bar b}_{j}^{(2)}} \ \ \xi_{Q_j} (I, \phi) \le  \pi_b^{(2)} \ \  \xi (I, \phi)
	$$
	where $\xi_{Q_j} (I, \phi)$ is a constant depending only upon $(I, \phi)$, as long as the service times are \richardCol{multiplied by} a constant. 
	\richardCol{So} $||\Theta_{lin}||_{1,2} \to 0$ as $\pi_b^{(2)} \to 0$ (see equations (\ref{Eqn_FixedC})-(\ref{Eqn_FixedU})). 
	
	Thus there exists a $\bar{\pi}_b^{(2)}$ such that $\Theta_{lin}$ is a contraction when ${\bar b}_{j}^{(2)} < \bar{\pi}_b^{(2)}$ for all $j$. \richardCol{  \emph{Therefore the contraction rate $\mu $ can be chosen strictly smaller than $1$, and independent of $\sigma$}, and:}
	$$
|| \Theta (\phi_c, \phi_u; \sigma) - \Theta (\phi_c', \phi_u'; \sigma) ||_{1, 2} \le \mu
||  (\phi_c, \phi_u) -  (\phi_c', \phi_u') ||_{1, 2} 
	\mbox{ for all } \sigma. 
	$$
	
	\richardCol{Because of stationarity, $\Theta$ has a unique fixed point for every $\sigma$ denoted $(C_*^{\sigma}, U_*^\sigma) = (C_*^{\sigma},U_{*,0}^{\sigma}, \cdots, U_{*,N-1}^{\sigma})$.} \richardCol{As discussed earlier we obtain the continuity of fixed points w.r.t. $\sigma$}. 


\underline {Continuity  of $\Theta (C^\sigma_*, U^\sigma_*; \sigma)$ with respect to $\sigma$:} 
It is difficult to obtain the continuity of $\Theta$ as $\sigma \to \infty$ at all   possible $(\phi_c, \phi_u)$. However we \richardCol{only} need  
continuity  at the fixed points $\{(C^\sigma_*, U^\sigma_*)\}$. \richardCol{These fixed points} additionally satisfy the equations (\ref{Eqn_C*sigmaUB})-(\ref{Eqn_C*sigmaUB1}) given in footnote \ref{Eqn_footnoteUB}. 

\noindent{\bf Step 1:} We first prove that ($\mbox{for any given  } \sigma' \le \infty$): 
\begin{eqnarray}
\label{Eqn_Reqd}
\lim_{\sigma \to \infty} 
||\Theta (C^{\sigma'}_*, U^{\sigma'}_*; \sigma)(q) -  \Theta (C^{\sigma'}_*, U^{\sigma'}_* ; \infty) (q) ||_{2}  = 0,
\mbox{ for all } q.
\end{eqnarray}
 From (\ref{Eqn_FixedC})-(\ref{Eqn_FixedU}), the difference 
$
\Theta (C_*^{\sigma'}, U_*^{\sigma'}; \sigma) -  \Theta (C_*^{\sigma'}, U_*^{ \sigma'}; \infty)$ \richardCol{involves} terms like $\varpi_{Q_j}$, 
defined as below\footnote{\label{Eqn_footnoteUB} From equation (\ref{Eqn_Cnq}),  $C_*^\sigma (q)$ is the sum of stationary quantities corresponding to 
$T_n (q)$ and $T_{n-1} (|{\cal C}|) - T_{n-1} (q)$ and is \richardCol{upper bounded by} $2 C_*^\sigma (|{\cal C}|)$ in distribution. This is true for all $q$, so that
\vspace{-5mm}
\begin{eqnarray}
\label{Eqn_C*sigmaUB}
C_*^\sigma (q)  \stackrel{distribution}{\le} 2C_*^\sigma (|{\cal C}|)  \mbox{ for all } q.
\end{eqnarray} 
\richardCol{Furthermore,  $|{\cal C}|$ or $0$ are arbitrary points of the polling system and so:}
\vspace{-1mm}
\begin{eqnarray}
\label{Eqn_C*sigmaUB1}
\frac{1}{2} C_*^\sigma (q')  \stackrel{distribution}{\le} C_*^\sigma (q)  \stackrel{distribution}{\le} 2C_*^\sigma (q')  \mbox{ for any } q, q'.
\end{eqnarray} 
}:
\begin{eqnarray*}
\varpi_{Q_j}^\sigma (q ) :=  {\cal }{\cal T}_{Q_j} ([\delta^\sigma(q), q), C_*^{\sigma'}) \stackrel{distribution}{\le} 2 {\cal }{\cal T}_{Q_j} (([\delta^\sigma(q), q), C_*^{\sigma'}(|{\cal C}|) ). 
\end{eqnarray*}  \richardCol{The r.h.s in the above equation denotes} the workload arriving in the interval $[\delta^\sigma(q), q)$ \richardCol{during time} $C_*^{\sigma'}(|{\cal C}|) $. \richardCol{By independence of the service times, for any interval $I$:}
\begin{eqnarray*}
E\left [ \left ( {\cal T}_{Q_j} (I, C_*^{\sigma'}(|{\cal C}|) ) \right )^2   \right ] \hspace{-30mm}&& \\
&=& E \left [ {\cal N}_{Q_j} (I, C_*^{\sigma'}(|{\cal C}|) \myCol{)} \left({\cal N}_{Q_j} (I, C_*^{\sigma'}(|{\cal C}|) \myCol{)} - 1 \right ) 
\left( E\left [B_{Q_j} (Q_j)| \{ Q_j \in I \} \right ] \right )^2 \right ] \\
&& + E \left [ {\cal N}_{Q_j} (I, C_*^{\sigma'}(|{\cal C}|) \myCol{)} 
E\left [ \left (B_{Q_j} (Q_j) \right )^2 |\{ Q_j \in I \} \right ] \right ]. \\
&\le & \sup_q \left ( b_{j}(q) \right )^2 \ \ E \left [ \left ({\cal N}_{Q_j} (I, C_*^{\sigma'}(|{\cal C}|) \myCol{)} \right )^2 \right]  \\
&& 
+ \left | \sup_q b_{j}^{(2)} (q) -  \inf_q \left ( b_{j} (q) \right )^2 \right |  E \left [ {\cal N}_{Q_j} (I, C_*^{\sigma'}(|{\cal C}|) \myCol{)} \right]. 
\end{eqnarray*}
The term ${\cal N}_{Q_0} (I, C_*^{\sigma'}(|{\cal C}|)\myCol{)}$ \richardCol{is} the number of Poisson arrivals that occurred in the interval $I$ for a duration of $C_*^{\sigma'}(|{\cal C}|)$ and its moments are:
\begin{eqnarray*}
E\left [ {\cal N}_{Q_0} (I, C_*^{\sigma'}(|{\cal C}|) \myCol{)}\right ] & =& \lambda P_{Q_0} ( I ) E[C_*^\sigma(|{\cal C}|)] \mbox{ and } \\
E\left [ \left ( {\cal N}_{Q_0} (I, C_*^{\sigma'}(|{\cal C}|) \myCol{)}\right )^2 \right ] &= &(\lambda P_{Q_0} ( I ))^2 E\left [ \left ( C_*^\sigma(|{\cal C}|) \right )^2 \right ] +
\lambda P_{Q_0} ( I ) E[C_*^\sigma(|{\cal C}|)].
\end{eqnarray*}
The term ${\cal N}_{Q_j} (I, C_*^{\sigma'}(|{\cal C}|) \myCol{)}$, with $j > 1$, \richardCol{is} the number of arrivals rerouted to interval $I$ out of the arrivals ${\cal N}_{Q_0} ([0,|{\cal C}|], C_*^{\sigma'}(|{\cal C}|)\myCol{)}$, whose moments can easily be calculated:
\begin{eqnarray*}
E\left [ {\cal N}_{Q_j} (I, C_*^{\sigma'}(|{\cal C}|) \myCol{)}\right ] & =& \lambda {\hat \epsilon}_j P_{Q_j} ( I ) E[C_*^\sigma(|{\cal C}|)] \mbox{ and } \\
E\left [ \left ( {\cal N}_{Q_j} (I, C_*^{\sigma'}(|{\cal C}|)\myCol{)} \right )^2 \right ] &= &(\lambda {\hat \epsilon}_j  P_{Q_j} ( I ))^2 
E\left [ \left (C_*^\sigma(|{\cal C}|)\right)^2 \right  ]
 +
\lambda {\hat \epsilon}_j  P_{Q_j} ( I ) E[C_*^\sigma(|{\cal C}|)].
\end{eqnarray*}
\richardCol{Thus, for all $q$, the difference  
$\Theta (C_*^{\sigma'}, U_*^{\sigma'}; \sigma) -  \Theta (C_*^{\sigma'}, U_*^{ \sigma'}; \infty)$ is a sum of terms multiplied by $P_{Q_j}([\delta^\sigma(q), q))= Prob (Q_j \in [\delta^\sigma(q), q))$ for some $0 \le j \le N$.}
  For all $q$, the half open intervals $[\delta^\sigma(q), q)$ converge to \richardCol{the} empty set and the probability  $P_{Q_j} (I)$ converges to zero as $\sigma \to \infty$. \richardCol{We have proven the result (\ref{Eqn_Reqd}).}

\noindent{\bf Step 2:} Convergence in $||.||_{1,2}$ norm.

  \richardCol{From the calculations of the previous step, the difference  $\Theta (C_*^{\sigma'}, U_*^{\sigma'}; \sigma) -  \Theta (C_*^{\sigma'}, U_*^{ \sigma'}; \infty)$  is upper bounded by a constant at all $q$.}
This and the point-wise convergence imply the following convergence using the BCT (for any $\sigma'$):
\begin{eqnarray}
|| \Theta (C_*^{\sigma'}, U_*^{\sigma'}; \sigma) -  \Theta (C_*^{\sigma'}, U_*^{ \sigma'}; \infty) ||_{1, 2}  \hspace{67mm} \nonumber \\
= E_{Q_0} \left [ ||\Theta (C^{\sigma'}_*, U^{\sigma'}_*; \sigma)(Q_0) -  \Theta (C^{\sigma'}_*, U^{\sigma'}_* ; \infty) (Q_0) ||_{2} \right ]
\to 0 \mbox{ as } \sigma \to \infty.  \hspace{3mm}
\label {EqnCntThetaVsC}
\end{eqnarray}

 \underline{$||.||_{1,2}$-Continuity  of fixed points with respect to $\sigma$:}

By the definition of the fixed points,
 \begin{eqnarray*}
 (C_*^\sigma, U_*^\sigma) -   (C_*^\infty, U_*^\infty)  &=&  \Theta (C_*^\sigma, U_*^\sigma; \sigma) -  \Theta (C_*^\infty, U_*^\infty; \infty)  \\
 && \hspace{-32mm} = \Theta (C_*^\infty, U_*^\infty; \sigma) -  \Theta (C_*^\infty, U_*^\infty; \infty) + \Theta (C_*^\sigma, U_*^\sigma; \sigma) -  \Theta (C_*^\infty, U_*^\infty; \sigma) .
 \end{eqnarray*}
 We recall that for any $\sigma$:
$$
 || \Theta (C_*^\sigma, U_*^\sigma; \sigma) -  \Theta (C_*^\infty, U_*^\infty; \sigma) ||_{1, 2} \le \mu  || (C_*^\sigma, U_*^\sigma ) -   (C_*^\infty, U_*^\infty ) ||_{1,2}
$$ 
 And hence,
 \begin{eqnarray*}
 || (C_*^\sigma, U_*^\sigma) -   (C_*^\infty, U_*^\infty ) ||_{1,2} &
 \le & \frac{1}{1-\mu}  || \Theta (C_*^\infty, U_*^\infty; \sigma) -  \Theta (C_*^\infty, U_*^\infty; \infty) ||_{1,2}  .
 \end{eqnarray*}
  Thus, using (\ref{EqnCntThetaVsC}),  as $\sigma \to \infty$ 
  \begin{eqnarray}
 || (C_*^\sigma, U_*^\sigma) -   (C_*^\infty, U_*^\infty) ||_{1,2} \to 0.
  \label{EqnCntFixedVsC}
 \end{eqnarray}

\underline{Almost sure ($Q_0$)-Continuity  of fixed points with respect to $\sigma$:}

The convergence in (\ref{EqnCntFixedVsC}) can be seen as  $L^1$ convergence of the ordered tuple, $\{ \left ( || C_*^\sigma (Q_0, .) ||_2,  || U_*^\sigma (Q_0, .) ||_2, \right ) \}$, which implies convergence in probability (with respect to measure $P_{Q_0}$). Then there exists a subsequence $\sigma_k \to \infty$ along which we have almost sure convergence. This implies the following almost sure convergence (note that $\left ( || C_*^\sigma (Q_0, .) ||_2 \right )^2 = E[ C_*^{\sigma} (Q_0, .)^2] = c_*^{\sigma 2} (Q_0)$ 
for any $Q_0$ and see (\ref{Eqn_ProdNorm})):
\begin{eqnarray}
\label{Eqn_ASCgnce}
P_{Q_0}  \left (  \left \{ Q_0: c_*^{\sigma_k 2} (Q_0)   \tends{\sigma_k}{\infty}   c_*^{\infty 2} (Q_0)  \right \} \right ) = 1. 
\end{eqnarray}
 By taking expectation on the left inequality of  (\ref{Eqn_C*sigmaUB1}) of footnote \ref{Eqn_footnoteUB} with $q' = |{\cal C}|$, 
$$
c_*^{\sigma_k 2}(|{\cal C}|) = E[ C_*^{\sigma_k} (|{\cal C}|, .)^2]  \le  2 E[ C_*^{\sigma_k} (q, .)^2]  = 2 c_*^{\sigma_k 2}(q)  \mbox{ for all } q$$ 
and this implies (since $P_{Q_0}$ is a probability measure)
 \begin{eqnarray}
 \label{Eqn_BndCyc0}
c_*^{\sigma_k 2}(|{\cal C}|)   \le  2 E_{Q_0}\left [  c_*^{\sigma_k 2}(Q_0) \right ].
 \end{eqnarray}
 \richardCol{As in footnote} \ref{Eqn_footnoteUB}, we also have $ c_*^{\sigma 2} (q) \le 2 c_*^{\sigma 2} (|{\cal C}|)$ for all $q.$
 Hence, using the BCT, almost sure convergence of (\ref{Eqn_ASCgnce}) implies: 
\begin{eqnarray*}
 E_{Q_0} \left [   c_*^{\sigma_k 2}(Q_0) \right ] \to  E_{Q_0} \left [  c_*^{\infty 2}(Q_0)\right ].
\end{eqnarray*}
Thus there exists a ${\bar \sigma} < \infty$ and a $\omega < \infty$ such that (see  (\ref{Eqn_BndCyc0})):
 \begin{eqnarray}
 \label{Eqn_BndCyc0Fin}
   c_*^{\sigma_k 2} (|{\cal C}|)  \le  2 \left ( E_{Q_0}\left [ c_*^{\infty 2} (Q_0)\right ] + \omega \right ) \mbox{ for all } \sigma_k > {\bar \sigma} .
 \end{eqnarray}
Using again the  inequality (\ref{Eqn_C*sigmaUB1}) of footnote \ref{Eqn_footnoteUB}, we have the following uniform upper bound:
\begin{eqnarray}
 \label{Eqn_BndCyc0Fin2}
  c_*^{\sigma_k 2} (q)  \le  4 \left ( E_{Q_0}\left [ E[ c_*^{\infty 2} (Q_0)] \right ] + \omega \right ) \mbox{ for all } \sigma_k > {\bar \sigma} 
  \mbox{ and for all  }  q. \ \ \ \ \Box
 \end{eqnarray}

\section*{Appendix B}

\begin{lemma}
\label{Lemma_TimDiff}
For every $q$ and $j$  (with  ${\hat N}^\sigma_j(y, q)$ defined recursively via (\ref{Eqn_NBar}), (\ref{Eqn_NHat}) and (\ref{Eqn_hatN0}) in the body of the proof),
{\small
\begin{eqnarray}
{ \nu}_j^\sigma(q) &=&   b_{{j}} (q)  E_{Q_{(j-1)}} \left [ {\hat N}_j^\sigma (Q_{(j-1)}, q )  \right ] \nonumber \\
&& \hspace{-15mm} +  \lambda \tau_*^\sigma (|{\cal C}|)  
\left ( {\hat \epsilon}_j \rho_{{j}} \left( \left  (q, \delta^\sigma(q)+ \frac{|{\cal C}|}{\sigma} \right ) \right ) +\sum_{j'=1}^{j-1}{\hat \epsilon}_{j'} \rho_{{j'}} \left( \left  [\delta^\sigma(q), \delta^\sigma(q)+ \frac{|{\cal C}|}{\sigma} \right ) \right )  \right ) \hspace{7mm}\label{Eqn_nuj}
\end{eqnarray} } Furthermore,
$E_{Q_j} \left[ \nu_j^\sigma(Q_j)  b_{{j}}(Q_j) \right ] \to
 E_{Q_j} \left [\nu_j^\infty (Q_j)  b_{{j}}(Q_j) \right ]$.
\end{lemma}
{\bf Proof :}
We call the user under consideration (who is $j$-rerouted to $q$ and for whom the time difference, $\nu_j^\sigma(q)$, is being calculated) the tagged user.  \richardCol{ $\nu_j^\sigma (q)$ is the amount of time between the instant at which the external service for users belonging to $[\delta^\sigma(q), \delta^\sigma(q)+|{\cal C}|/\sigma)$ starts and the instant at which the tagged user's service starts (which happens before the 'external' service).}
\richardCol{The time difference $\nu_j^\sigma(q)$ includes the time taken to serve}: 
\begin{enumerate}
\item the $j$-rerouted users belonging to \Alter{strip}{interval}
$(q, \delta^\sigma(q)+|{\cal C}|/\sigma)$
(arrival position order service);  
\item all the $j'$-rerouted users of \Alter{strip}{interval} $[\delta^\sigma(q), \delta^\sigma(q)+|{\cal C}|/\sigma)$ with $1 \le j' < j$  (users with maximum completed services are served first);  and  
\item $j$-rerouted users that arrived exactly at $q$ and after the tagged user (FIFO).
\end{enumerate}
\noindent All 3 points exist in a discrete polling system while the first point does not exist in the \cts polling system. For a \cts system, $\nu_j^\sigma$  is non zero only where $P_{Q_j}$ has atoms, i.e. at $\{q_{j,i}\}_{i \le M_j}$. 
(see (\ref{Eqn_Arrivalprobs})).   

 By Lemma \ref{LemmaRe} \Col{(see equation (\ref{Eqn_nuj}) of the statement of the lemma)}, the average time taken to serve the $j$-rerouted users belonging to \Alter{strip}{interval}
$(q, \delta^\sigma(q)+|{\cal C}|/\sigma)$ and all the $j'$-rerouted users with $1 \le j' < j$, is given by the last two terms of
${ \nu}_j^\sigma$ defined above. 

\richardCol{We now consider     $j$-rerouted users of point 3.} We calculate this quantity conditioned on the location (say $y$) at which the $j$-th service was received (whose distribution equals $P_{Q_{(j-1)}}$ by independence) by the tagged user. 
 Let ${\hat N}^\sigma_j (y, q)$ represent the total number of users $j$-rerouted to point $q$ after the tagged user, who received his $j$-th service at $y$. These will be among the ones that  were $(j-1)$-rerouted to\footnote{
$\epsilon_j P_{Q_{j}} (\{q\})$ fraction of these users \richardCol{are contained in} ${\hat N}^\sigma_j (y, q).$ Note that the users of points (i)-(iii) are the ones that were
$(j-1)$-served  after the tagged user and if they are  $j$-rerouted to the same point $q$ then by FIFO order will receive the $(j+1)$-service also after the tagged user.}: 

(i) segment $(y, \delta^\sigma(q))$   (if $y < \delta^\sigma(q)$);  

(ii)  segment $(y, |{\cal C}|) \cup [0, \delta^\sigma(q))$  (if $y \ge  \delta^\sigma (q)$);

(iii)  point $\{y\}$ but after the tagged user. 

\Col{Let ${\hat {\cal T}}_{Q_{(j-1)}}$ \richardCol{denote}  ${\cal T}_{Q_{(j-1)}}$ with service times replaced by $1$, as done in the proof of Lemma \ref{LemmaRe}.} {
Then the average number of  $(j-1)$-rerouted users of case (i) equals $E[{\hat {\cal T}}_{Q_{(j-1)}} ((y,\delta^\sigma(q)), C).$}
By  Lemma \ref{LemmaRe},  the average of the sum due to cases (i) and (ii)  \richardCol{is} \footnote{In case (ii) we consider users rerouted from two \richardCol{successive} cycles. By stationarity, the two cycles are equal in distribution.}:
\begin{eqnarray}
\label{Eqn_NBar}
{\bar N}^\sigma_j (y, q) &:=&   \lambda \tau_*^\sigma (|{\cal C}|) {\hat \epsilon}_{j-1}
\left ( P_{Q_{(j-1)}} ((y,  \delta^\sigma(q))) 1_{\{ y < \delta^\sigma(q)\}}  \right . \nonumber \\ &&  \hspace{30mm} \left.  + P_{Q_{(j-1)}} (( \delta^\sigma(q), y))  1_{\{  \delta^\sigma(q) \le y\}}  \right ) . \hspace{19mm}
\end{eqnarray} 
The average number of users of case (iii) \richardCol{is} calculated by induction. The users $j$-rerouted to $q$ from \richardCol{point} $y$ after the tagged user must be  among the users that were $(j-1)$-rerouted to $y$ after the tagged user in the previous cycle, because of FIFO service. Hence,
{
\begin{eqnarray}
\label{Eqn_NHat}
{\hat N}^\sigma_j (y, q) =  \left\{ \begin{array}{lllll}
  \epsilon_j P_{Q_j} (\{q\})    \left ( {\bar N}^\sigma_j (y, q) 
+ E_{Q_{(j-2)}} \left [{\hat N}^\sigma_{(j-1)} (Q_{(j-2)}, y) \right ] \right )  &\mbox{if}  &  j \ge 2 \\
  \epsilon_1 P_{Q_1} (\{q\})    \left ( {\bar N}^\sigma_1 (y, q) 
+ {\hat N}^\sigma_{0}( y)  \right )  &\mbox{if} &  j = 1 \\
\end{array} \right.  \nonumber \\
\end{eqnarray}}
\richardCol{with ${\hat N}_0^\sigma (q)$ the external users arriving at point $q$ before the tagged user, (during his residual 
cycle time). ${\hat N}_0^\sigma (q)$ is calculated as in (\ref{Eqn_UserdueRes}):}
  \begin{eqnarray}
  \label{Eqn_hatN0}
  {\hat N}_0^\sigma ( q) =   \sum_{i=1}^{M_0} p_{0,i} 1_{\{q_{0,i}\}} (q) E[C_R^\sigma(q)]  \mbox{ for all } y.
  \end{eqnarray}
  \richardCol{By integrating with respect to $P_{Q_{(j-1)}}$} we obtain the average number of $j$-rerouted users to point $q$ after the tagged user. Applying Wald's lemma (by independence) gives the time required to \richardCol{serve these users}. The first part of the lemma is proven. When $\sigma = \infty$ the same quantities define $\{\nu^\infty_j \}$  for \cts systems.  Here the  terms with empty sets are deleted, and some sets are replaced by singletons etc. \richardCol{For instance the third term is deleted in (\ref{Eqn_nuj}) and the fourth term is replaced by $\rho_{{j'}} (q)$).}

The convergence part of the lemma follows by Lemma \ref{Lemma_fI}, Theorems  \ref{Thrm_FirstMoments}, \ref{Thrm_SecondMoments} and BCT 
recursively starting with $j=1$ up to $j = N-1$. $\Box$

\Col{
\begin{lemma}
\label{Lemma_fI} 
Let $g$ be a  nonnegative function with $||g||_\infty < \infty$ and define $f(I) := \int_I g(y) P_{Q_j} (dy)$ for every  subset, $I \subset [0, |{\cal C}|]$.  Then, 
\begin{eqnarray*} 
\mbox{ {\bf i)} for any decreasing sequence $\{I^\sigma\}_\sigma$, }  \ \ f(I^\sigma) \to  f( \cap_\sigma I^\sigma)    \hspace{2mm} \mbox{ and so }  \hspace{-20mm} &\\
 \mbox{ {\bf ii)} if $  \cap_\sigma I^\sigma = \{q\}$,  then } f(I^\sigma) \to  P_{Q_j} (\{q\}) g(q);   \hspace{15mm}
&\hspace{-20mm} \mbox{ {\bf iii) } if   $\cap_\sigma I^\sigma = \emptyset,$ then }  f(I^\sigma) \to 0.  
  \end{eqnarray*}
\end{lemma}}
{\bf Proof :} The measure $g(y) P_{Q_j} (dy)$ is a finite measure and this lemma follows by continuity of probability measures, i.e. that 
$Prob ( \cap_n B_n) = \lim_n  Prob (B_n)$.  $\Box$

\section*{Appendix C: Little's law and Palm moments for stationary queueing systems}

	Little's law is a well known result (\cite{Whitt}) in queueing theory linking the expected waiting time of a typical user with the expected number of users in the system. We recall here a version of Little's law derived as in \cite{Palm}. In particular ergodicity of the input process is not required, stationarity is sufficient.
We also  derive a relationship between  the Palm \Cut{distribution of inter-arrival times and their stationary distribution} \Col{stationary moments and the stationary moments}, mentioned in \myCut{footnote} \myCol{subsection}
\ref{footnote_palm}. 
 We use independent notations from the body of the article.
 
	\subsection*{Palm probability}
	Consider $(T_n,W_n)_{n \in \ZZ}$ a stationary marked point process on $\RR$ with marks taking value in a measurable space ${\cal K}$. We use the convention $T_0 \leq 0 < T_1$. Write $\lambda$ its intensity defined as:
\begin{equation}
	\lambda = \frac{1}{t_2 - t_1}  E[\sum_{n \in \ZZ} \indic_{[t_1,t_2)}(T_n)].
\end{equation}
with arbitrary $t_1 < t_2$. \mynCol{The definition is independent of $t_1$ and $t_2$, because}  we consider a stationary process. Consider  \Col{$K \subset {\cal K}$} a measurable set. The \emph{Palm probability} of $K$ is defined as:
\begin{equation}\label{eq:palm_def}
	  P^{0}[ W_0 \in K] = \frac{E\left[\sum_{n \in \ZZ} \indic_{ [t_1,t_2) \times K}(T_n,W_n)\right ]}{E\left[\sum_{n \in \ZZ} \indic_{ [t_1,t_2)}(T_n) \right ]}
 = \frac{E\left[\sum_{n \in \ZZ} \indic_{[t_1,t_2) \times K}(T_n,W_n) \right ]}{\lambda(t_2 - t_1)} .
\end{equation}	
	\Col{One can easily verify that $P^0 [T_0 = 0] = 1$ (let $t_1 = 0$ in the above definition), so that this probability is defined conditioned on the event that the first arrival occurs at a reference time i.e. at $0$.}

	Consider a measurable function $f:\RR \times K \to \RR$, we have the Campbell-Mecke-Little formula:
\begin{equation}\label{eq:campbell_mecke}
	E[\sum_{n \in \ZZ} f(T_n,W_n)] = \lambda \int_{\RR \times {\cal K}} f(t,w) dt P^{0}[W_0 \in dw].
\end{equation}
	  For $f(t,w) = \indic_{[t_1,t_2) \times K}(t,w)$, \eqref{eq:campbell_mecke} is a reformulation of the definition \eqref{eq:palm_def}. By linearity, \eqref{eq:campbell_mecke} also holds for a simple function i.e a linear combination of indicator functions. Finally, \eqref{eq:campbell_mecke} can be extended to all measurable functions by a monotone class argument.
	
\subsection*{Little's law}
	Little's law is a special case of \eqref{eq:campbell_mecke}. Consider a queueing system with $T_n$ the arrival time of the $n$-th user and $W_n$ the time he stays in the system. The $n$-th user exits the system at time $T_n + W_n$. \mynCol{With ${\cal K} = \mathbb{R}$ and choosing}  \newline $f(t,w) = \indic_{(-\infty,0] \times [0,+\infty)}(t,t+w)$, equation \eqref{eq:campbell_mecke} becomes: 
\begin{equation}\label{eq:campbell_little}
	E[ \sum_{n \in \ZZ} \indic_{(-\infty,0] \times [0,+\infty)}(T_n,T_n+W_n)] =  \lambda \int_{\cal K} w P^{0}[ W_0 \in dw] = \lambda E^{0}[W_0].
\end{equation}
	It is noted that $\indic_{(-\infty,0] \times [0,+\infty)}(T_n,T_n+W_n) = 1$ means that the $n$-th user enters the system before time $0$ (or at time $0$), and that he leaves the system after time $0$ (or at time $0$). Namely, the $n$-th user is in the system at time $0$. This is Little's law: the expected number of users present in the system at time $0$ (or any other time by stationarity) equals the arrival rate multiplied by the time spent in the system by a typical user.
	
	It is noted that \eqref{eq:campbell_little} is valid \emph{without} assuming ergodicity of the input process, only stationarity is required.

	In this article, we apply Little's law to the users waiting to be served, \emph{excluding the user currently receiving service}. This is also valid by applying the same reasoning as above and defining $W_n$ to be the time the $n$-th customer waits until he starts receiving service.

	\subsection*{Palm moments}
	The relation between Palm moments and stationary moments can be derived using again the
	equation ~\eqref{eq:campbell_mecke}.  Consider now $f(t,w) = \indic_{(-\infty,0] \times [0,+\infty)}(t,t+w)$, and $W_n = T_{n+1} - T_n$. Then~\eqref{eq:campbell_mecke} proves that:
\begin{equation}\label{eq:palm_moment_1}
	1 = \lambda E^{0}[T_1 - T_0],
\end{equation}
similarly, for $f(t,w) = w \indic_{(-\infty,0] \times [0,+\infty)}(t,t+w)$, and $W_n = T_{n+1} - T_n$:
\begin{equation}\label{eq:palm_moment_2}
\expec{T_1 - T_0} = \frac{\lambda}{2} E^{0}[(T_1 - T_0)^2].
\end{equation}
	Combining \eqref{eq:palm_moment_1} and \eqref{eq:palm_moment_2} gives the required relation between Palm moments and the stationary moments:
\begin{equation}\label{eq:palm_moment_3}
\expec{T_1 - T_0} = \frac{E^{0}[(T_1 - T_0)^2]}{2 E^{0}[T_1 - T_0]}.
\end{equation}




\section*{Appendix D: Notations index}

We recall the notations used in this article. Not all notations used in the proofs is listed for clarity. The notations specific to the Ferry WLAN and waste collector of Sections~\ref{section_fwlan} and~\ref{section_collect} are listed separately. Quantities which are random are denoted r.v (random variable).

\begin{table}
	\centering
\begin{tabular}{|ll|}
	\hline
	General & \\
	\hline	
	$E[.]$ & mathematical expectation\\
	$E^0[.]$ & Palm expectation\\
	$E_R[.]$ & expectation with respect to random variable $R$\\
	$Prob(.)$ & probability\\
	${\cal U}(.)$ & uniform distribution \\
	${\cal C} $ & circle on which the server moves\\ 
	$\alpha$ & server speed \\	
	$\lambda$ & arrival rate of external users\\	
	$N < +\infty$ & maximal number of reroutings \\
	$0 \leq j \leq N$ & index for re-routings, \\
										& $0$-th rerouting denotes an external arrival\\
	$\epsilon_j$ & probability of being re-routed \\
	& after receiving the $j-1$-th service\\
	$\hat{\epsilon}_j$ & probability of going through $j$ re-routings\\ 
	\mynCol{$\check {\epsilon}_j^k := \Pi_{i=j}^{k} \epsilon_i$}  & \mynCol{probability of going through $k$-reroutings} \\ & \mynCol{conditional to having gone through $j$ reroutings.} \\
	\hline
	Arrival locations & \\
	\hline
	$Q_j \in {\cal C}$ & location of arrival for the $j$-th rerouting (r.v)\\
	$q \in {\cal C}$ & index for the position of arrival\\ 
	$P_{Q_j}(dq)$ & probability distribution of $Q_j$\\
	$M_j$ & number of atoms of $P_{Q_j}$\\
	$q_{j,i} \in {\cal C} , 1 \leq i \leq M_j$ & location of the $i$-th atom of $P_{Q_j}$\\
	$p_{j,i} \in {\cal C} , 1 \leq i \leq M_j$ & measure of the $i$-th atom of $P_{Q_j}$\\
	$f_{Q_j}(q)dq$ & non-atomic part of $P_{Q_j}$\\
	\hline
	Service times & \\
	\hline
	$B_{Q_j}$ & service time after the $j$-th rerouting (r.v)\\
	$\overline{b}_{j} ,\overline{b}_{j}^{(2)}$ & first and second moment of $B_{Q_j}$\\
	$b_{j}(q) ,b_{j}^{(2)} (q)$ & first and second moment of $B_{Q_j}$,\\
	    &  conditioned on $Q_j=q$\\	
	$\bar{\pi}_b^{(2)} > 0 $ & threshold on the second moments \\ 
	& of the service times\\
	\hline
	Virtual workload & \\
	\hline
	$V$ & expected stationary virtual workload\\
	$V_{rrt}$ & $V$ with rerouting\\
	$V_{rrt}^{sym}$ & $V_{rrt}$ under symmetric conditions (see \ref{subsec:special_cases})\\
	$V_{g}$ & $V$ with gated service\\
	$V_{gg}$ & $V$ with globally gated service\\
	$V_{g}^{sym}$ & $V_{g}$ under symmetric conditions\\
	\hline
	\end{tabular}
\end{table}
	
	\begin{table}
	\centering
\begin{tabular}{|ll|}
	\hline
	Discretization & \\
	\hline
	$\sigma \in \NN$ & number of discretization levels\\
	$\delta^{\sigma}(q) \in {\cal C}$ & point where the server serves customers waiting at $q$ \\
	& in the discretized system\\
	$\delta^{\infty}(q)$ & identity map\\
	$V_{rrt}^{\sigma}$ & $V_{rrt}$ for the discretized system\\
	\hline
	Ferry WLAN & (specific to Section~\ref{section_fwlan}) \\
	\hline
	$\Delta = [-D_1,D_1] \times [-D_2,D_2]$ & area in which the ferry moves\\
	$D_1 > 0$ & half-length of $\Delta$\\
	$D_2 > 0$ & half-breadth of $\Delta$\\
	${\cal C}_{d_1,d_2}$ & ferry path determined by distances $(d_1,d_2)$ \\
	& from the edges of $\Delta $\\
	${\cal C}_{d}$ & ${\cal C}_{d,d}$\\
	$x \in \Delta$ & location of a user\\
	$B$ & service time of users (r.v)\\
	$B(x)$ & service time of users located at $x$ (r.v)\\
	$\overline{b} ,\overline{b}^{(2)}$ & first and second moment of $B$\\
	$q(x) \in {\cal C}$ & point of the ferry path at which the ferry   \\
	& serves users located at $x$ \\
  $b(q) ,b^{(2)}(q)$ & first and second moment of $B$ \\
	& for users located at $x$, with $q = q(x)$\\
	$d$ & parameter of the ferry path\\
	$V_{fwlan}(d)$ & $V$ for the ferry WLAN for $d$\\
	$V_{fwlan}^{hybrid}$ & $V$ for the ferry WLAN and hybrid service\\
	$d^*$ & value of $d$ optimizing $V_{fwlan}$\\
	$d_b^*$ & value of $d$ optimizing $\overline{b}$\\
	$_d$ , $_u$ & subscript for downlink/uplink\\
	$I(q) \subset \Delta$ & points of $\Delta$ served when the ferry is at $q$\\
	$\eta > 0$ & number of bytes downloaded by a user\\
	$X,Y \in \Delta$ & positions of sources\\
	 & and destinations (\acp{r.v})\\
	$P_X,P_Y$ & probability distributions of $X,Y$\\
	$\Psi$ & arrival measure of requests on the ferry path\\
	$E_{\Psi}$ & expectation with respect to $\Psi$\\
 	$q_{i} \in {\cal C} , 0 \leq i \leq 3$ & location of the $i$-th atom of $\Psi$\\
	$p_{i} \in [0,1] , 0 \leq i \leq 3$ & measure of the $i$-th atom of $\Psi$\\
	$f_{\psi}(q)dq$ & non-atomic part of $\Psi$\\
	$\vartheta_{_P}$ & transmit power\\
	$\beta$ & path-loss exponent\\
	$\kappa(d)$ & achievable data rate at distance $d$ from the ferry\\
	
	\hline
\end{tabular}
\end{table}

	\begin{table}
	\centering
\begin{tabular}{|ll|}
	\hline
	Waste collector & (specific to Section~\ref{section_collect}) \\
	\hline
	$p$ & time needed for picking up the goods\\
	$d$ & time needed for discarding the goods\\
	$f_Q$ & arrival measure\\
	$x_d$ & point where goods are discarded\\
	$V_{WC}$ & $V$ for waste collector\\
	$x_d^*$ & value of $x_d$ optimizing $V_{WC}$\\
	\hline
\end{tabular}
\end{table}

\end{document}